%% file: IJSR_preparedness_combined.tex
\documentclass[pdflatex,iicol,sn-basic]{sn-jnl}

\usepackage[utf8]{inputenc}
\DeclareUnicodeCharacter{00D7}{\ensuremath{\times}}
\usepackage[T1]{fontenc}
\usepackage{amsmath,amssymb,bm}
\usepackage{graphicx}
\usepackage{booktabs}
\usepackage{tabularx}
\usepackage{longtable}
\usepackage{array}
\usepackage{multirow}
\usepackage{enumitem}
\usepackage{microtype}
\usepackage{subcaption}
\usepackage[table]{xcolor}
\usepackage[section]{placeins}
\usepackage{pdflscape}
\usepackage{adjustbox}
\usepackage{cuted}

\definecolor{rowblue}{RGB}{243,240,235}
\definecolor{betweenblue}{RGB}{238,246,252}
\definecolor{interactionblue}{RGB}{228,240,248}

\newcolumntype{P}[1]{>{\raggedright\arraybackslash}p{#1}}
\newcolumntype{Y}{>{\raggedright\arraybackslash}X}

\graphicspath{{figures/}{scr_figures/}{scr_supplement/}{scr_v10/}{scr_v9/}{scr_v8/}{scr_v7/}}
\newcommand{\fitab}[1]{\begin{adjustbox}{max width=\textwidth}#1\end{adjustbox}}

\newcommand{\Ready}{\mathrm{Ready}}
\newcommand{\Sig}{\mathrm{Sig}}
\newcommand{\Comp}{\mathrm{Comp}}
\newcommand{\Gap}{\mathrm{Gap}}

\begin{document}

\title[Ready for What?]{Ready for What? Rethinking AI and Robotics Preparedness for Adoption and Policy}

\author*[1]{\fnm{Peng} \sur{Wang}}
\email{peng.wang@surrey.ac.uk}

\author[2]{\fnm{Naomi} \sur{Adel}}

\author[3]{\fnm{Amy} E. \sur{Morgan}}

\author[4]{\fnm{Folayo} \sur{Aina}}

\author[5]{\fnm{Demos} \sur{Parapanos}}

\author[2]{\fnm{Vikas} \sur{Mackevicius}}

\author[4]{\fnm{Teslim Olayiwola} \sur{Salahudeen}}

\affil*[1]{
\orgdiv{Centre for Vision, Speech and Signal Processing (CVSSP)},
\orgname{University of Surrey},
\orgaddress{\city{Guildford}, \country{UK}}
}

\affil[2]{
\orgname{Manchester Metropolitan University},
\orgaddress{\city{Manchester}, \country{UK}}
}

\affil[3]{
\orgdiv{Biomedical Research \& Innovation Centre}
\orgname{University of Salford},
\orgaddress{\city{Manchester}, \country{UK}}
}

\affil[4]{
\orgname{University of Lancashire},
\orgaddress{\city{Preston}, \country{UK}}
}

\affil[5]{
\orgname{University of Cumbria},
\orgaddress{\city{Ambleside}, \country{UK}}
}

\abstract{Efforts to accelerate AI and robotics adoption require evidence about where communities are prepared to act and where support is still needed. However, conclusions based mainly on average differences between people or stakeholder groups can obscure relationships that emerge when the same person evaluates different challenges. We address this problem using a repeated card-based survey in which 982 participants provided 15,200 evaluations of 17 AI and robotics challenges. Each challenge was rated on common 1-5 measures of significance, complexity and readiness, where readiness refers to perceived community preparedness and available resources rather than personal competence or realised adoption. Because each participant evaluated multiple challenges, the design separates stable differences between respondents from challenge-specific deviations within the same respondent. This distinction materially changes the interpretation of preparedness. Within the same respondent, a challenge rated one point more complex than their usual level is associated with approximately 0.21 points lower readiness ($p<0.001$). By contrast, respondents who generally rate challenges as more complex do not systematically report lower readiness ($p=0.29$). Significance is positively associated with readiness, while unusually high complexity modestly weakens this challenge-specific alignment. These relationships also vary strongly across challenge families, and professional background remains associated with adjusted preparedness assessments. On applied cards, confidence, trust and related perceptions provide substantial additional information about readiness, including for held-out participants. For policymakers and organisations, the findings show that averaging across stakeholders can hide challenge-specific barriers. Readiness assessments should therefore preserve both differences between stakeholder groups and variation within the same stakeholders across challenges. Effective adoption and literacy strategies should ask not only \emph{who} appears ready, but \emph{which challenges} they find unusually difficult and whether the likely constraint concerns implementation, capability, assurance or resources.}


\maketitle

\section{Introduction}

Artificial intelligence (AI) and robotics are moving from isolated pilots towards wider use in organisational decision making, service delivery, production, healthcare and knowledge work. As adoption expands, the policy question is no longer simply whether an AI-enabled application is technically feasible or potentially valuable. An equally important question is whether the people, organisations, resources and institutions surrounding its use are sufficiently prepared to support implementation. This has led to growing interest in AI readiness as a socio-technical rather than purely technological problem.

Existing research has substantially advanced this understanding. Organisational AI readiness frameworks show that successful adoption depends on combinations of strategic alignment, data, knowledge, technical resources, organisational culture and management commitment \citep{NewJohnk2021,NewHolmstrom2022}. A recent systematic review similarly identifies technological, organisational and environmental conditions as important components of AI readiness \citep{NewAliKhan2025}. Sector-specific studies reinforce the importance of context. Readiness to adopt AI is shaped by the organisational environment in which implementation takes place \citep{NewHradecky2022}, while recent healthcare research shows that adoption depends on how value, capability and organisational support are developed and recognised across different actors \citep{NewDuus2026}. Together, these studies establish that readiness is multidimensional and context dependent.

A related body of work has examined technology acceptance and individual willingness to use new technologies. The Technology Acceptance Model, for example, identify perceived usefulness, ease of use, performance expectancy, effort expectancy, social influence and facilitating conditions as important predictors of behavioural intention and technology use \citep{Davis1989,Venkatesh2003,Venkatesh2012}. The Technology Readiness Index instead captures a person's broader propensity to embrace new technologies through enabling and inhibiting dispositions \citep{Parasuraman2000}. These approaches provide important explanations of whether individuals are willing or inclined to adopt technology. However, willingness to use a technology is conceptually different from judging whether the wider community has the resources, knowledge, infrastructure and institutional support required to address a particular AI or robotics challenge. An individual may be willing to use an AI system while believing that the surrounding community is poorly prepared to govern, integrate or support it. Conversely, substantial organisational capability may exist even when an individual remains reluctant to use the technology.

The literature also increasingly shows that attitudes towards AI depend on the specific application being considered. Horowitz et al. \citep{NewHorowitz2024}, for example, show that familiarity and expertise do not lead to uniform attitudes across autonomous vehicles, surgery, weapons and cyber defence. Familiarity can increase acceptance in some contexts while producing greater scepticism in others. A recent work points in the same direction. It shows that 1) trust in organisational AI changes as developers, managers and users accumulate different forms of knowledge and experience; 2) responsible AI concerns in tourism are influenced by ambiguity, anxiety and trustworthiness requirements 3) healthcare adoption depends on the organisational context in which AI value is created and recognised; and 4) the effects of embedded generative AI can differ according to user skill and task complexity \citep{NewSong2026}. These studies suggest that AI is not a single adoption object. The same broad technology can create different benefits, risks and implementation demands depending on the challenge, application and stakeholder involved.

Perceived complexity is particularly important within this context. Existing AI readiness frameworks recognise technical and organisational complexity as potential barriers to adoption \citep{NewJohnk2021,NewAliKhan2025}. Industry evidence on trustworthy AI similarly identifies practical difficulties involving data quality, accountability, transparency, regulatory compliance, technological robustness and human oversight \citep{NewMcCormack2026}. However, much of this evidence treats complexity at the level of the technology, organisation or implementation programme. Less is known about whether the relationship operates at the level of a specific challenge as perceived by the same stakeholder. This distinction matters because two different processes can produce an observed negative relationship between complexity and readiness. Some people may simply tend to rate most challenges as difficult and most communities as poorly prepared. Alternatively, the same person may identify particular challenges as unusually difficult relative to the other challenges they consider, and those same challenges may also be judged as less prepared for. These are substantively different explanations and imply different policy responses.

Trust and confidence provide another important part of the current evidence base. Trust in AI is multidimensional and context sensitive, involving beliefs about competence, reliability, safety, fairness, transparency and appropriate reliance \citep{NewAfroogh2024,NewMehrotra2024,NewAquilino2025}. The objective is therefore not simply to maximise trust, but to achieve appropriately calibrated trust: insufficient trust may inhibit useful adoption, while excessive trust can create inappropriate reliance \citep{NewMehrotra2024}. Trust may also change as stakeholders gain experience or receive new information about AI capabilities \citep{NewDaly2025,NewNikolova2025}. Confidence is related but different, reflecting how capable or comfortable a person feels when engaging with a technology. These perceptions may therefore help explain why apparently similar AI challenges receive different preparedness assessments, but they should not be treated as equivalent to readiness itself.

Stakeholder heterogeneity adds a further complication. AI implementation involves people with different professional roles, experience, responsibilities and exposure to risk. Classic technology adoption models recognise age, experience and social context as potential moderators of technology use \citep{Venkatesh2003,Venkatesh2012}, while AI-specific studies show that professional role, familiarity and task characteristics can influence trust, support and behavioural responses \citep{NewHorowitz2024,NewDaly2025,NewSong2026}. A software developer, engineer, manager, researcher and non-technical user may therefore assess the same technology from very different positions. These differences are potentially important for policy because a lower preparedness assessment from one professional group does not necessarily indicate lower competence or greater resistance. It may instead reflect greater exposure to implementation, validation, workflow, governance or resource constraints.

Despite these advances, three important limitations remain. First, much of the existing readiness literature assesses individuals, organisations or sectors using aggregate or cross-sectional measures. Such approaches are useful for describing overall differences, but they can conceal relationships that occur within the same person across different AI and robotics challenges. In particular, a relationship observed when the same stakeholder compares different challenges may disappear when responses are aggregated and different people are compared with one another. This is important for policy because averaging across stakeholders can obscure challenge-specific barriers.

Second, existing studies typically examine importance or value, implementation difficulty, readiness, trust and related perceptions in separate analytical traditions. Technology acceptance research focuses heavily on value and willingness to use; readiness research focuses on organisational capability and implementation conditions; and human-AI interaction research frequently focuses on trust, confidence and risk. There is comparatively little evidence showing how perceived significance, complexity and preparedness coexist across multiple concrete AI and robotics challenges evaluated by the same respondents.

Third, stakeholder heterogeneity is often represented through demographic controls or average group comparisons. This can identify whether professional groups differ overall, but it does not show whether the same stakeholder changes their assessment depending on the challenge being considered. As a result, aggregated comparisons may miss the interaction between the person evaluating the technology and the particular challenge they are evaluating.

\begin{figure*}[btp]
\centering

\begin{subfigure}[t]{0.19\textwidth}
    \centering
    \includegraphics[
        width=\linewidth,
        height=4.2cm,
        keepaspectratio
    ]{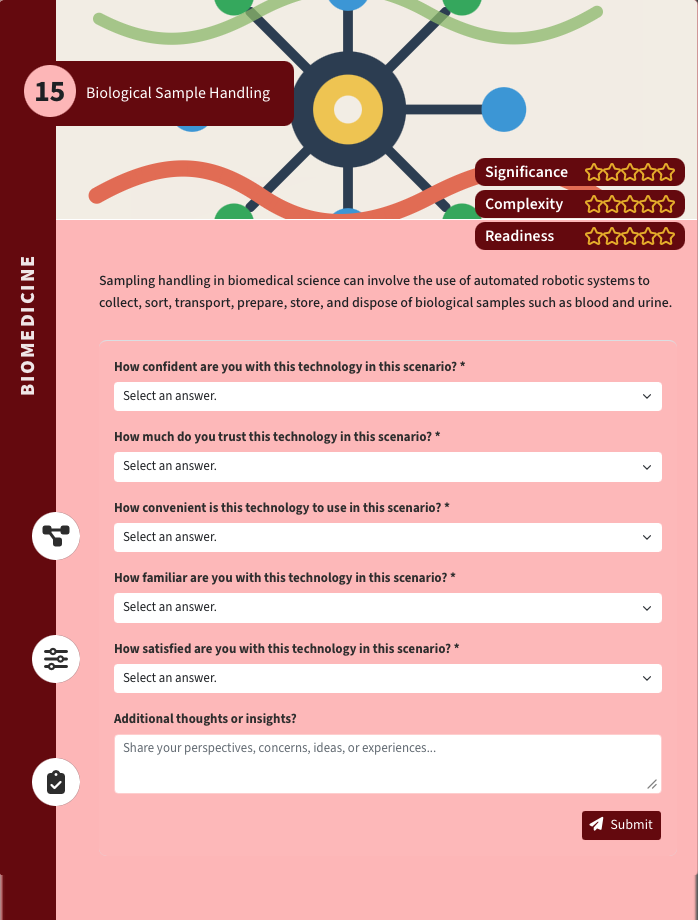}
    \caption{Biomedical}
    \label{fig:panel_a}
\end{subfigure}
\hfill
\begin{subfigure}[t]{0.19\textwidth}
    \centering
    \includegraphics[
        width=\linewidth,
        height=4.2cm,
        keepaspectratio
    ]{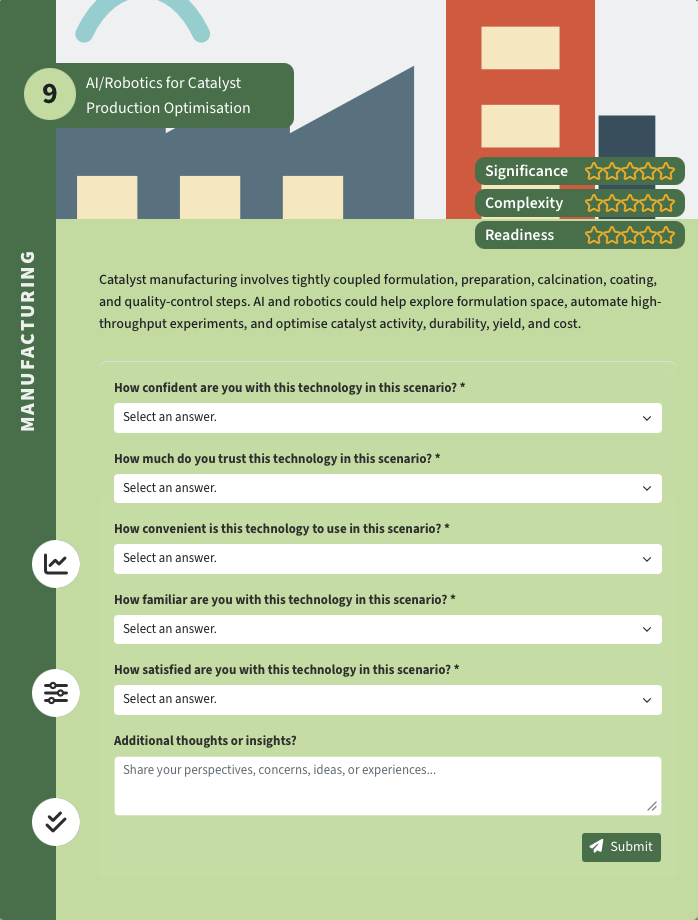}
    \caption{Manufacturing}
    \label{fig:panel_b}
\end{subfigure}
\hfill
\begin{subfigure}[t]{0.19\textwidth}
    \centering
    \includegraphics[
        width=\linewidth,
        height=4.2cm,
        keepaspectratio
    ]{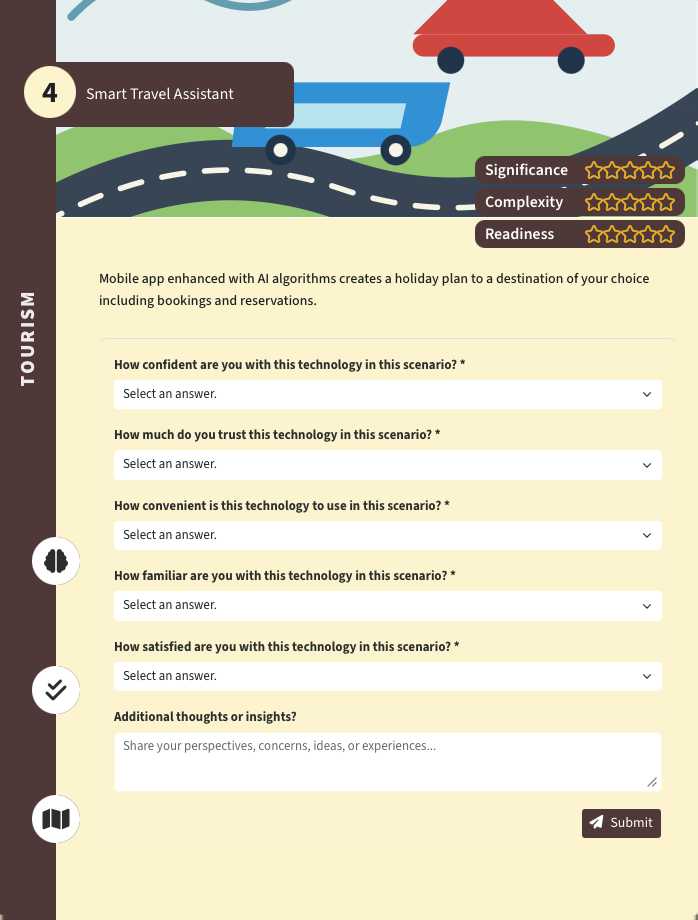}
    \caption{Hospitality}
    \label{fig:panel_c}
\end{subfigure}
\hfill
\begin{subfigure}[t]{0.19\textwidth}
    \centering
    \includegraphics[
        width=\linewidth,
        height=4.2cm,
        keepaspectratio
    ]{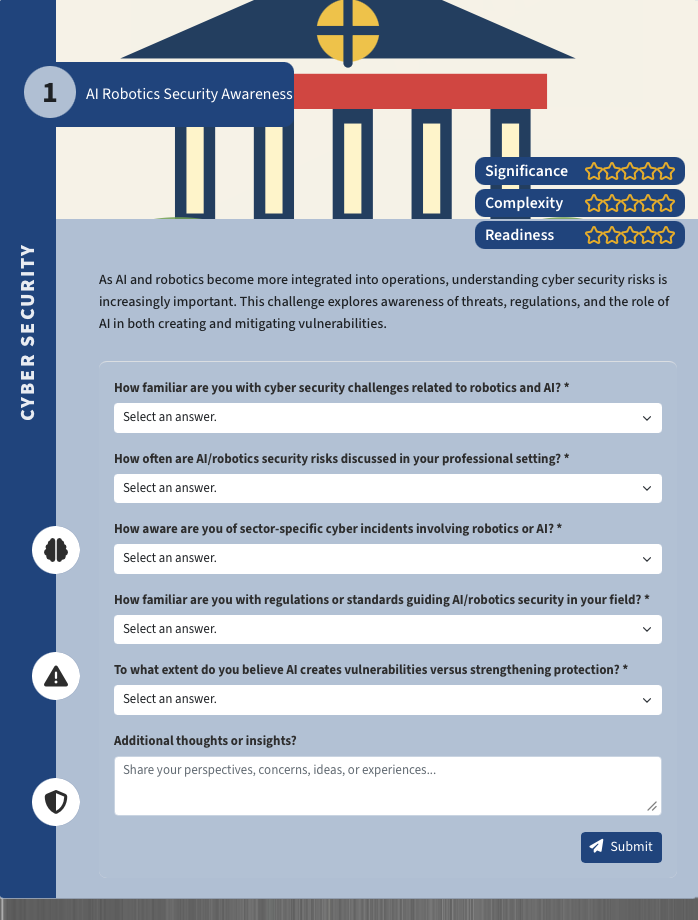}
    \caption{Cyber Security}
    \label{fig:panel_d}
\end{subfigure}
\hfill
\begin{subfigure}[t]{0.19\textwidth}
    \centering
    \includegraphics[
        width=\linewidth,
        height=4.2cm,
        keepaspectratio
    ]{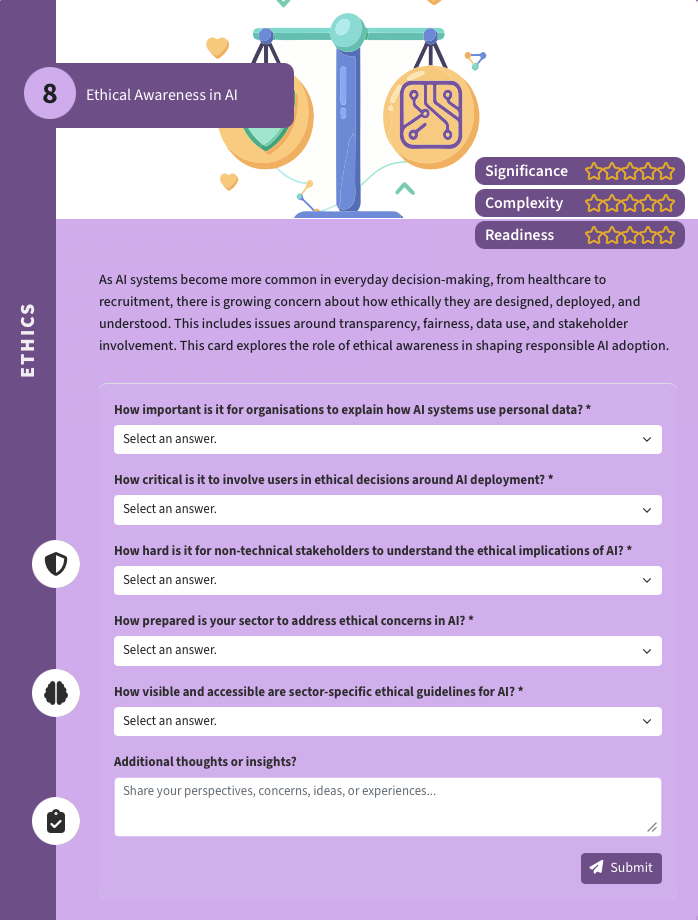}
    \caption{Ethics}
    \label{fig:panel_e}
\end{subfigure}

\caption{
Examples of challenge cards from the five AI and robotics challenge families:
(a) Biomedical, (b) Manufacturing, (c) Hospitality,
(d) Cyber Security, and (e) Ethical Awareness.
}
\label{fig:five_families}
\end{figure*}

This study addresses these limitations using a repeated card-based survey in which 982 participants provided 15,200 evaluations of 17 AI and robotics challenges, comprising three Biomedical, six Manufacturing, four Hospitality, three Cyber Security and one Ethical Awareness card. Every challenge used the same three core ratings: significance, complexity and readiness. Significance captures how much impact respondents believe addressing a challenge could have; complexity captures its perceived technical, organisational and societal difficulty; and readiness captures perceived community preparedness and available resources. The 13 Biomedical, Manufacturing and Hospitality cards, referred to as `applied cards', additionally measured confidence, trust, familiarity, convenience and satisfaction.

The repeated design is central to the study. Because each participant evaluates multiple challenges, we can distinguish two forms of information that would otherwise be mixed together. The between-person component captures whether some respondents generally give higher or lower ratings across the challenges they evaluate. The within-person component captures whether a particular challenge is rated higher or lower than that same respondent's usual level. This allows us to ask not only whether people who generally perceive greater complexity also report lower readiness, but whether a specific challenge that appears unusually complex to the same person is also perceived as less prepared for. This distinction is particularly important for understanding what may be lost when stakeholder responses are aggregated.

The study makes three contributions. First, it provides a challenge-level account of perceived AI and robotics preparedness by examining significance, complexity and readiness jointly across multiple concrete challenges rather than relying only on general assessments of AI readiness. Second, it separates within-person challenge-specific relationships from between-person differences, allowing us to test whether important preparedness relationships remain visible when individual baselines are taken into account. This provides direct evidence on whether aggregation across people can conceal challenge-specific barriers. Third, it examines how preparedness assessments differ across challenge families and professional groups, while confidence, trust and related perceptions provide additional diagnostic information about possible capability, assurance and implementation constraints.

These contributions are intended to support a more targeted approach to AI and robotics adoption and literacy. For policymakers and organisations, the relevant question is not only whether a workforce, sector or community appears ready on average. It is also which challenges particular stakeholders perceive as unusually difficult, where preparedness assessments differ across professional groups, and whether low preparedness is associated with implementation difficulty, confidence, trust or other resource constraints. Preserving these differences can provide more useful evidence for deciding where adoption support, skills development, assurance mechanisms and targeted interventions should be tested.

Accordingly, we ask five research questions: \textit{RQ1} how significance, complexity and readiness vary across AI and robotics challenges; \textit{RQ2} which challenges exhibit the largest descriptive significance-readiness gaps and how those gaps relate to complexity; \textit{RQ3} how challenge-specific significance and complexity relate to readiness once stable between-person differences are separated; \textit{RQ4} how preparedness assessments and complexity sensitivity vary across professional and demographic groups; and \textit{RQ5} whether confidence, trust and related perceptions provide additional diagnostic information on the applied cards.

\begin{figure*}[ptb]
\centering
\includegraphics[width=\textwidth]{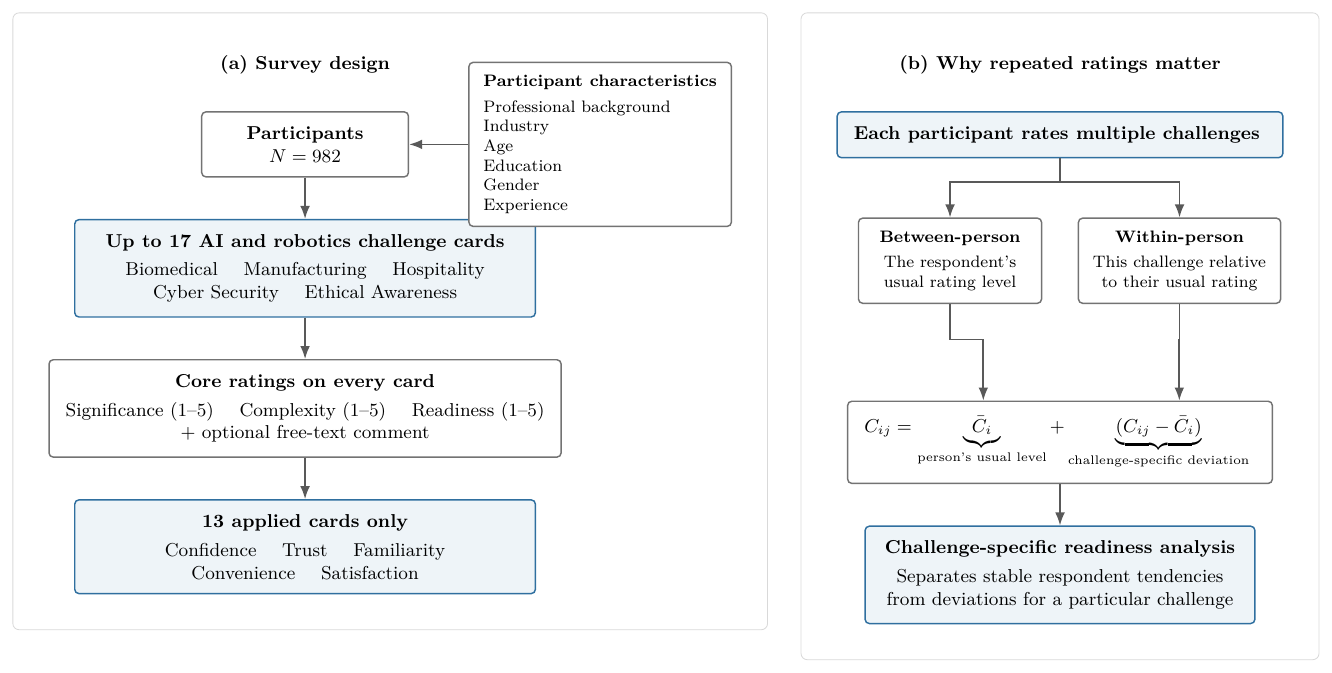}
\caption{Card-based repeated-measures survey design. Each participant evaluates multiple AI and robotics challenges using common significance, complexity and readiness ratings. Applied cards additionally collect confidence, trust and related perceptions.}
\label{fig:design}
\end{figure*}

\begin{table*}[ptb]
\centering
\caption{Operational definitions and analytical roles of the three core survey constructs and the derived significance-readiness gap.}
\label{tab:constructs}
\scriptsize
\renewcommand{\arraystretch}{0.95}
\rowcolors{2}{white}{rowblue}
\begin{tabularx}{\textwidth}{P{1.8cm}P{6.5cm}Y}
\toprule
Construct & Operational definition & Analytical role \\
\midrule
Significance
&
1-5 survey rating of the potential impact of improving awareness of, or addressing, the challenge
(1 = very low impact; 5 = critical impact).
&
Indicates how much impact respondents believe addressing the challenge could have.
\\
Complexity
&
1-5 survey rating of the combined technical, organisational and societal difficulty of raising awareness of, or solving, the challenge
(1 = very simple; 5 = very complex), considering barriers, coordination, societal factors and resource requirements.
&
Indicates how difficult respondents perceive the challenge to be.
\\
Readiness
&
1-5 survey rating of current community preparedness and available resources to address the challenge and promote awareness
(1 = very poor readiness; 5 = excellent readiness), considering initiatives, educational resources, research progress and community engagement.
&
Primary outcome: perceived community preparedness and available resources for the challenge.
\\
Gap $(S-R)$
&
Derived analytical measure, not a survey item. Calculated as significance minus readiness for the same participant-card evaluation.
&
Descriptive measure of the difference between perceived significance and readiness; not the primary regression outcome.
\\
\bottomrule
\end{tabularx}
\rowcolors{2}{white}{white}
\renewcommand{\arraystretch}{1.0}
\end{table*}

\section{Data and methods}

\subsection{Survey design and constructs}

Participants completed an online card-based survey covering 17 AI and robotics challenges across five challenge families: Biomedical, Manufacturing, Hospitality, Cyber Security and Ethical Awareness. Examples from each family are shown in Figure~\ref{fig:five_families}. Each card used the same three core ratings, i.e., significance, complexity and readiness, on a five-point scale.

We use `applied cards' to refer to the 13 Biomedical, Manufacturing and Hospitality cards. These cards additionally used the same measures of confidence, trust, familiarity, convenience and satisfaction, together with optional free text. The three Cyber Security cards and one Ethical Awareness card used different specialist items and are therefore excluded from analyses requiring these additional common measures.

The use of a common rating structure across multiple concrete challenges supports direct comparison between cards and, because participants evaluated multiple cards, provides repeated observations for separating respondent-level tendencies from challenge-specific variation. We used one question for each rating because participants assessed up to 17 challenges. This kept the survey manageable and allowed the same questions to be compared across challenges. The ratings therefore represent participants’ direct assessments of each specific challenge, rather than comprehensive measures of every aspect of significance, complexity or readiness. Figure~\ref{fig:design} summarises the survey structure.

Table~\ref{tab:constructs} reports the operational definitions of the three core survey ratings. Readiness refers specifically to the respondent's judgement of community preparedness and available resources for the challenge presented. It should not be interpreted as personal competence, behavioural intention, realised adoption or independently audited organisational capability. The survey did not define a fixed institutional boundary for `community', but the contextual information on each card would bound the referent challenge-specific. In this paper we use both `readiness' and `perceived preparedness' to refer to this judgement.
Because the same gap can arise from different rating combinations, such as $(\Sig,\Ready)=(5,4)$ and $(2,1)$, interpretation is therefore based on the underlying ratings rather than the gap alone.

Confidence, trust, familiarity, convenience and satisfaction are analysed as additional challenge-specific perceptions on the 13 applied cards. They are kept separate from readiness because they measure respondents' perceptions of the technology rather than community preparedness.

The significance-readiness gap, $S-R$, was derived for the analysis and was not asked directly in the survey. It is used descriptively to indicate whether significance is rated above or below readiness for the same participant-card evaluation. It is not treated as a target that readiness should equal significance. For the $j$-th card completed by participant $i$, the significance-readiness gap is defined as
\begin{equation}
\Gap_{ij}=\Sig_{ij}-\Ready_{ij}.
\label{eq:gap}
\end{equation}

\subsection{Sample}

Participants were recruited through Prolific. Of 1,235 registered participants, 240 provided no card responses. Among the 995 participants with card data, attention-check exclusions yielded an analytical sample of 982 participants contributing 15,200 usable card evaluations; 864 participants completed all 17 cards. Table~\ref{tab:sample} summarises the main participant profile. Complete industry, education and experience distributions are reported in Supplementary Table~S1.

Because each participant contributed multiple card evaluations, the 15,200 observations are not independent. We therefore first estimated a null participant random-intercept model,
\begin{equation}
R_{ij}=\mu+u_i+\varepsilon_{ij},
\end{equation}
\noindent
where $R_{ij}$ is the readiness rating given by participant $i$ to challenge $j$, $\mu$ is the overall mean readiness level, $u_i$ captures stable differences in readiness ratings between participants, and $\varepsilon_{ij}$ captures variation among individual card responses that is not explained by the participant-specific intercept. The estimated variance components were $\hat{\sigma}_{u}^{2}=0.442$ and $\hat{\sigma}_{e}^{2}=0.919$, giving an intraclass correlation of
\begin{equation}
\mathrm{ICC}=
\frac{\hat{\sigma}_{u}^{2}}
{\hat{\sigma}_{u}^{2}+\hat{\sigma}_{e}^{2}}
=0.325.
\end{equation}

\noindent
This tells us that 32.5\% of the variation in readiness reflects stable differences between participants, while the remaining 67.5\% occurs among individual card responses, including variation across challenges within the same participant and other residual variation (Supplementary Table~S6). This confirms that responses from the same participant are related and provides empirical justification for separating respondent-level tendencies from challenge-specific variation in the subsequent analysis.

\begin{table*}[pbt]
\centering
\caption{Participant profile for the analytical sample ($N=982$).}
\label{tab:sample}

\scriptsize
\renewcommand{\arraystretch}{0.95}
\setlength{\tabcolsep}{3pt}

\begin{minipage}[t]{0.27\textwidth}
\vspace{0pt}

\textbf{Gender}

\vspace{2pt}

\rowcolors{2}{white}{rowblue}
\begin{tabularx}{\linewidth}{@{}Xrr@{}}
\toprule
 & $n$ & \% \\
\midrule
Female & 494 & 50.3 \\
Male & 477 & 48.6 \\
Non-binary & 8 & 0.8 \\
Prefer not to say & 3 & 0.3 \\
\bottomrule
\end{tabularx}
\rowcolors{2}{white}{white}

\vspace{8pt}

\textbf{Age}

\vspace{2pt}

\rowcolors{2}{white}{rowblue}
\begin{tabularx}{\linewidth}{@{}Xrr@{}}
\toprule
 & $n$ & \% \\
\midrule
18--24 & 134 & 13.6 \\
25--34 & 360 & 36.7 \\
35--44 & 272 & 27.7 \\
45--54 & 123 & 12.5 \\
55--64 & 76 & 7.7 \\
65+ & 17 & 1.7 \\
\bottomrule
\end{tabularx}
\rowcolors{2}{white}{white}

\end{minipage}
\hfill
\begin{minipage}[t]{0.38\textwidth}
\vspace{0pt}

\textbf{Professional background}

\vspace{2pt}

\rowcolors{2}{white}{rowblue}
\begin{tabularx}{\linewidth}{@{}Xrr@{}}
\toprule
 & $n$ & \% \\
\midrule
Non-technical & 288 & 29.3 \\
Business & 163 & 16.6 \\
Other & 117 & 11.9 \\
Research & 76 & 7.7 \\
Operations & 73 & 7.4 \\
Engineering & 53 & 5.4 \\
Computer science & 49 & 5.0 \\
Data science & 48 & 4.9 \\
Software development & 46 & 4.7 \\
AI/ML & 38 & 3.9 \\
Design & 31 & 3.2 \\
\bottomrule
\end{tabularx}
\rowcolors{2}{white}{white}

\end{minipage}
\hfill
\begin{minipage}[t]{0.29\textwidth}
\vspace{0pt}

\textbf{Education}

\vspace{2pt}

\rowcolors{2}{white}{rowblue}
\begin{tabularx}{\linewidth}{@{}Xrr@{}}
\toprule
 & $n$ & \% \\
\midrule
Bachelor's & 415 & 42.3 \\
Master's & 233 & 23.7 \\
Some college & 160 & 16.3 \\
High school & 91 & 9.3 \\
Other & 83 & 8.5 \\
\bottomrule
\end{tabularx}
\rowcolors{2}{white}{white}

\vspace{8pt}

\textbf{Industry}

\vspace{2pt}

\rowcolors{2}{white}{rowblue}
\begin{tabularx}{\linewidth}{@{}Xrr@{}}
\toprule
 & $n$ & \% \\
\midrule
Technology & 136 & 13.8 \\
Other & 130 & 13.2 \\
Education & 105 & 10.7 \\
Healthcare & 101 & 10.3 \\
Retail & 100 & 10.2 \\
Remaining & 410 & 41.8 \\
\bottomrule
\end{tabularx}
\rowcolors{2}{white}{white}

\vspace{8pt}

\textbf{Experience}

\vspace{2pt}

\rowcolors{2}{white}{rowblue}
\begin{tabularx}{\linewidth}{@{}Xrr@{}}
\toprule
 & $n$ & \% \\
\midrule
Reported & 728 & 74.1 \\
Missing & 254 & 25.9 \\
\bottomrule
\end{tabularx}
\rowcolors{2}{white}{white}

\end{minipage}

\renewcommand{\arraystretch}{1.0}
\end{table*}

\subsection{Analytical strategy}

The analytical strategy moves from descriptive differences across challenges to challenge-specific associations within respondents, and then to heterogeneity across challenge families and stakeholder groups. We begin by characterising the 17-card landscape using card-level means for significance, complexity and readiness, together with the descriptive significance-readiness gap defined in Eq.~\eqref{eq:gap}. For brevity, we also use $S$, $C$ and $R$ to represent significance, complexity and readiness when necessary. Spearman correlations among $S$, $C$ and $R$ are used as exploratory summaries and are reported in Supplementary Table~S2. 

We next examine how much readiness varies across challenges for the same respondent. This analysis uses the 864 participants who completed all 17 cards and quantifies variation in readiness across a common challenge set. Its purpose is to assess whether a single respondent-level average would conceal substantial challenge-specific differences.

The primary inferential analysis separates respondents' general rating tendencies from challenge-specific deviations using a within--between, or person-mean-centred, decomposition. This distinction is important because the same observed rating can arise for different reasons: a respondent may generally assign high values across most challenges, or they may regard one particular challenge as unusually high relative to their own typical assessment. The decomposition allows these two sources of variation to be estimated separately. For significance,

\begin{equation}
\Sig_{ij}
=
\underbrace{\bar{\Sig}_i}_{\text{between-person}}
+
\underbrace{\left(\Sig_{ij}-\bar{\Sig}_i\right)}_{\text{within-person}},
\label{eq:sigdecomp}
\end{equation}

\noindent
and the same decomposition is applied to complexity. We therefore define

\begin{align}
\Sig^{pm}_{i}&=\bar{\Sig}_i, &
\Sig^{w}_{ij}&=\Sig_{ij}-\bar{\Sig}_i,\\
\Comp^{pm}_{i}&=\bar{\Comp}_i, &
\Comp^{w}_{ij}&=\Comp_{ij}-\bar{\Comp}_i.
\label{eq:wbdefs}
\end{align}

The person-mean terms, $\Sig^{pm}_{i}$ and $\Comp^{pm}_{i}$, capture between-person differences in respondents' usual rating levels. For example, $\Comp^{pm}_{i}$ asks whether respondents who generally perceive the challenge set as more complex also tend to report systematically different readiness levels. These person means may reflect differences in experience, attitudes, expectations or use of the response scale and are therefore interpreted as respondent-level associations rather than as fixed personal traits.

The within-person terms, $\Sig^{w}_{ij}$ and $\Comp^{w}_{ij}$, capture how a particular challenge differs from that respondent's own usual assessment. A positive $\Comp^{w}_{ij}$ means that respondent $i$ considers challenge $j$ more complex than the challenges they typically rate, whereas a negative value means that they consider it less complex than usual. The coefficient $\beta_{CW}$ therefore asks whether readiness differs when the same respondent perceives one challenge as more complex than their own average. Similarly, $\beta_{SW}$ asks whether readiness differs when the same respondent perceives one challenge as more significant than usual. For example, if a respondent's average significance rating across completed cards is 3.7 and they assign a significance rating of 5 to one challenge, then $\Sig^{pm}_{i}=3.7$ and $\Sig^{w}_{ij}=+1.3$. The decomposition therefore distinguishes the fact that this respondent generally gives relatively high significance ratings from the fact that this particular challenge is unusually significant for them.

\begin{table*}[ptb]
\centering
\caption{Within- and between-person terms used in the repeated-measures models. }
\label{tab:wbterms}
\footnotesize
\renewcommand{\arraystretch}{0.88}
\setlength{\tabcolsep}{3.5pt}

\begin{tabularx}{\textwidth}{@{}P{2.0cm}P{3.0cm}Y@{}}
\toprule
Term & Definition & Interpretation \\
\midrule

\rowcolor{rowblue}
$\Sig^{pm}_{i}$ & Mean significance across respondent $i$'s cards & Between-person: does generally higher significance correspond to different readiness? \\

$\Sig^{w}_{ij}$ & $\Sig_{ij}-\bar{\Sig}_i$ & Within-person: is this challenge more significant than usual for this respondent, and is readiness different? \\

\rowcolor{rowblue}
$\Comp^{pm}_{i}$ & Mean complexity across respondent $i$'s cards & Between-person: does generally higher complexity correspond to different readiness? \\

$\Comp^{w}_{ij}$ & $\Comp_{ij}-\bar{\Comp}_i$ & Within-person: is this challenge more complex than usual for this respondent, and is readiness different? \\

\rowcolor{rowblue}
$\Sig^{w}_{ij}\Comp^{w}_{ij}$ & Product of within-person deviations & Does complexity modify the within-person significance--readiness association? \\

\bottomrule
\end{tabularx}

\renewcommand{\arraystretch}{1.0}
\end{table*}

The interaction $\Sig^{w}_{ij}\Comp^{w}_{ij}$ tests whether the within-person association between significance and readiness changes when the same challenge is also perceived as unusually complex. It does not imply that objectively increasing the complexity of a challenge would cause readiness to change. Readiness remains the card-level outcome because the inferential target is each respondent's assessment of community preparedness for each specific challenge.

\subsubsection{Primary model}

The primary model is

\begin{strip}
\begin{equation}
\Ready_{ij}
=
\beta_0
+\beta_{SW}\Sig^{w}_{ij}
+\beta_{SB}\Sig^{pm}_{i}
+\beta_{CW}\Comp^{w}_{ij}
+\beta_{CB}\Comp^{pm}_{i}
+\beta_{SC}\Sig^{w}_{ij}\Comp^{w}_{ij}
+\boldsymbol{\alpha}^{\top}\mathbf{D}_j
+\boldsymbol{\gamma}^{\top}\mathbf{Z}_i
+\varepsilon_{ij}.
\label{eq:core}
\end{equation}
\end{strip}

\noindent
where $\mathbf{D}_j$ contains challenge-family indicators and $\mathbf{Z}_i$ contains professional background, age, education and gender. Manufacturing and non-technical are used as the reference categories for challenge family and professional background, respectively. Because the choice of reference category affects individual coefficient contrasts but not fitted values or omnibus tests, challenge-family and professional-background results are interpreted primarily using omnibus tests and adjusted marginal means across all groups.

The model is estimated using all card-level observations with complete significance, complexity and readiness ratings and the required covariates. Standard errors are clustered by participant to account for dependence among repeated ratings from the same respondent. Missing card responses are not imputed. Because some participants completed fewer than 17 cards, a complete-deck analysis using only participants who rated all 17 challenges is used as a robustness check to assess whether the person-mean estimates depend on the subset of cards completed.

Industry and experience are not included in the primary specification. Industry is examined in complementary adjusted analyses because it contains many categories, while experience is analysed separately using complete cases because experience information is missing for 25.9\% of participants.

\subsubsection{Challenge family and stakeholder heterogeneity}

We next test whether the within-person significance and complexity associations differ across challenge families and professional backgrounds. Omnibus Wald tests are conducted before subgroup estimates are interpreted. These tests assess the overall association of professional background with readiness, the interaction between within-person complexity and professional background, and the interactions of within-person significance and complexity with challenge family. Other demographic interactions are also explored with results reported where necessary.

The stakeholder analysis addresses two conceptually different questions. \textit{The first} asks whether professional groups differ in their overall assessments of community readiness after challenge characteristics and other covariates are taken into account. \textit{The second} asks whether the association between challenge-specific complexity and readiness itself differs across professional groups. These questions are analysed separately because a professional group may report systematically higher or lower readiness without necessarily showing a different response to perceived complexity.

\subsubsection{Applied card extension}

For the 13 Biomedical, Manufacturing and Hospitality cards that use equivalent additional questions, we examine whether confidence, trust, familiarity, convenience and satisfaction provide information beyond the core significance and complexity measures. A base model containing significance, complexity, challenge family and participant controls is compared with an extended model that additionally includes respondent-mean and within-person components of these perceptions.

Incremental model performance is assessed using three measures of explained variation. Adjusted $R^2$ indicates how much variation in readiness is explained by the model while accounting for the number of predictors included. Partial $R^2$ measures the additional explanatory contribution of the confidence, trust, familiarity, convenience and satisfaction variables beyond the base model. We also report participant-level five-fold cross-validated $R^2$ to assess how well the model explains readiness for respondents who were not used during model fitting. Cross-validation folds are defined by participant rather than by individual card response so that evaluations from the same respondent cannot appear in both the training and test sets. The cross-validated $R^2$ therefore provides evidence of generalisation to previously unseen respondents, but does not test whether the model generalises to new challenge cards.

Within Biomedical, Manufacturing and Hospitality, we additionally compare the conditional within-person coefficients for confidence and trust. Cyber Security and Ethical Awareness are excluded because equivalent confidence and trust measures were not collected for those cards. These challenge-family-specific comparisons are secondary analyses and are interpreted cautiously. Confidence, trust and the other additional perceptions are treated as diagnostic correlates rather than as causal mechanisms. Optional free-text analysis is also explored, with comment-selection criteria, sample sizes, sentiment scoring and theme-tagging procedures reported in Supplementary Methods~S-FT.

\subsubsection{Robustness checks}

We assess the robustness of the core within-person findings in three complementary ways. First, Eq.~\eqref{eq:core} is re-estimated using only the 864 participants who completed all 17 cards, ensuring that participant means are calculated from the same challenge set for everyone. Second, we estimate a participant and card fixed-effects model,

\begin{strip}
\begin{equation}
\Ready_{ij}
=
\beta_{SW}\Sig^{w}_{ij}
+\beta_{CW}\Comp^{w}_{ij}
+\beta_{SC}\Sig^{w}_{ij}\Comp^{w}_{ij}
+\alpha_i
+\delta_j
+\varepsilon_{ij},
\label{eq:twowayfe}
\end{equation}
\end{strip}

\noindent
where $\alpha_i$ absorbs all stable respondent-level differences and $\delta_j$ absorbs all stable card-specific differences. Because every participant saw the cards in the same fixed order, card identity and presentation position cannot be separated; $\delta_j$ therefore captures stable card/position differences. This specification provides a more demanding test of the within-person associations by asking whether they remain after accounting simultaneously for who the respondent is and which specific challenge is being rated.

Because readiness is measured on an ordered five-point scale, we estimate a proportional odds ordered logit and an ordinal Generalised Estimating Equation (GEE) with exchangeable within-participant correlation. These models report odds ratios for $\Sig^{w}$, $\Comp^{w}$ and $\Sig^{w}\times\Comp^{w}$ and assess whether the core conclusions depend on approximating readiness as a continuous outcome in the primary linear model. Additional pooled Ordinary Least Squares (OLS) and crossed linear mixed models are provided in the Supplementary Material. Model comparison focuses on whether the direction and substantive interpretation of the core within-person associations remain stable across specifications rather than on selecting the model that produces the most favourable goodness-of-fit statistic or $p$-value.

\section{Results}

\subsection{Preparedness varies across challenges and within respondents}

\begin{figure*}[btp]
\centering
\begin{subfigure}[t]{0.32\textwidth}
\centering
\includegraphics[width=\linewidth]{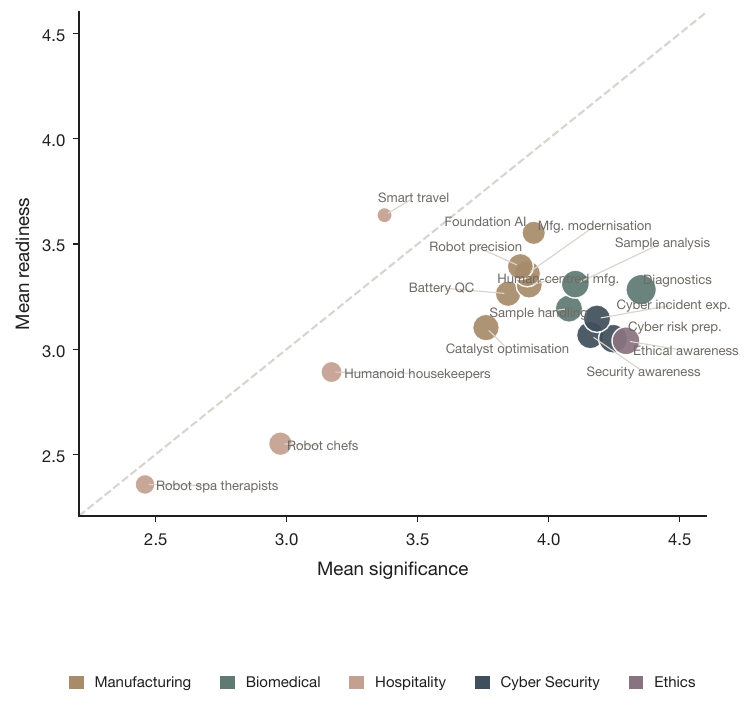}
\caption{Significance vs readiness}
\label{fig:landscape_a}
\end{subfigure}\hfill
\begin{subfigure}[t]{0.32\textwidth}
\centering
\includegraphics[width=\linewidth]{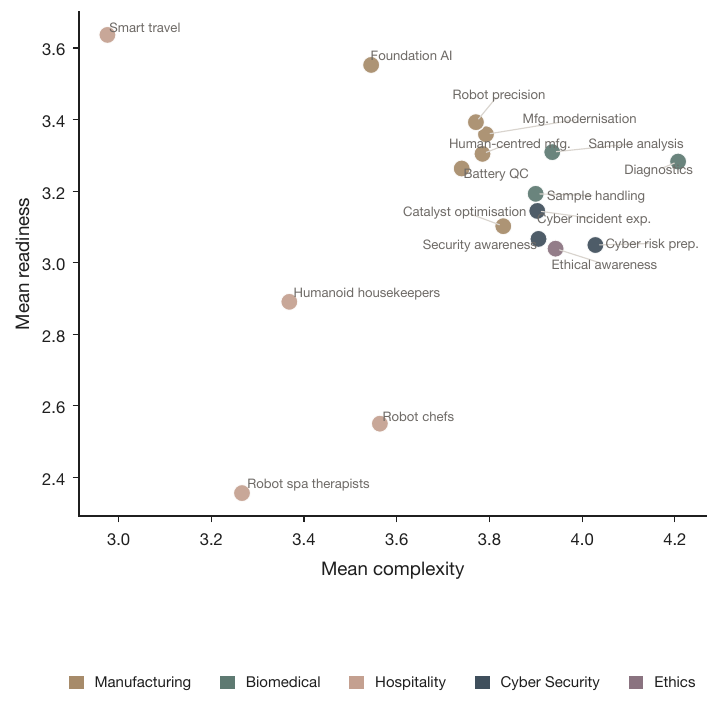}
\caption{Complexity vs readiness}
\label{fig:landscape_b}
\end{subfigure}\hfill
\begin{subfigure}[t]{0.32\textwidth}
\centering
\includegraphics[width=\linewidth]{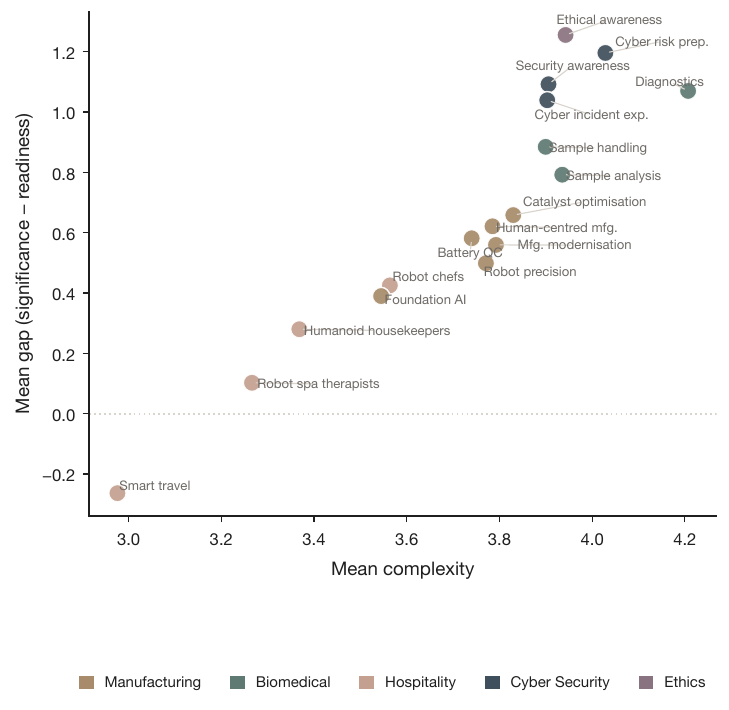}
\caption{Complexity vs gap}
\label{fig:landscape_c}
\end{subfigure}
\caption{Challenge-level SCR landscape. (a)~Mean significance versus mean readiness (point size $\propto$ mean complexity; dashed line $S=R$). (b)~Mean complexity versus mean readiness. (c)~Mean complexity versus the descriptive significance-readiness gap ($S-R$).}
\label{fig:landscape}
\end{figure*}

Figure~\ref{fig:landscape} summarises the 17-challenge SCR landscape. Figure~\ref{fig:landscape_a} compares mean significance with mean readiness, using the line $S=R$ only as a visual reference. Figure~\ref{fig:landscape_b}  shows mean complexity against mean readiness, while Figure~\ref{fig:landscape_c} shows mean complexity against the significance-readiness gap. The gap is used only as a descriptive measure; the main statistical analysis models readiness directly and includes significance and complexity as separate variables. The largest significance-readiness gaps are concentrated in Ethical Awareness, Cyber Security and Biomedical challenges, whereas several Manufacturing and Hospitality cards show much smaller differences. Smart Travel is the only challenge for which mean readiness slightly exceeds mean significance. This indicates that the mismatch between perceived importance and preparedness is strongly challenge-dependent rather than uniform across AI and robotics. Challenge-family residual readiness after adjusting for significance, complexity and professional background is summarised in Supplementary Fig.~S1 (with accompanying means in Supplementary Table~S6b).

\begin{table*}[btp]
\centering
\caption{Mean significance, complexity, readiness and significance-readiness gap across the 17 challenge cards.}
\label{tab:cards}
\scriptsize
\input{scr_tables_card.tex}
\end{table*}

Table~\ref{tab:cards} reports the exact challenge-level means. The results show that the significance-readiness gap should not be interpreted alone, because similar gaps can arise from different combinations of significance, complexity and readiness. We therefore interpret the three ratings jointly rather than rank challenges by the gap alone.

\begin{figure*}[tbp]
\centering
\begin{subfigure}[t]{0.58\textwidth}
\centering
\includegraphics[width=\linewidth]{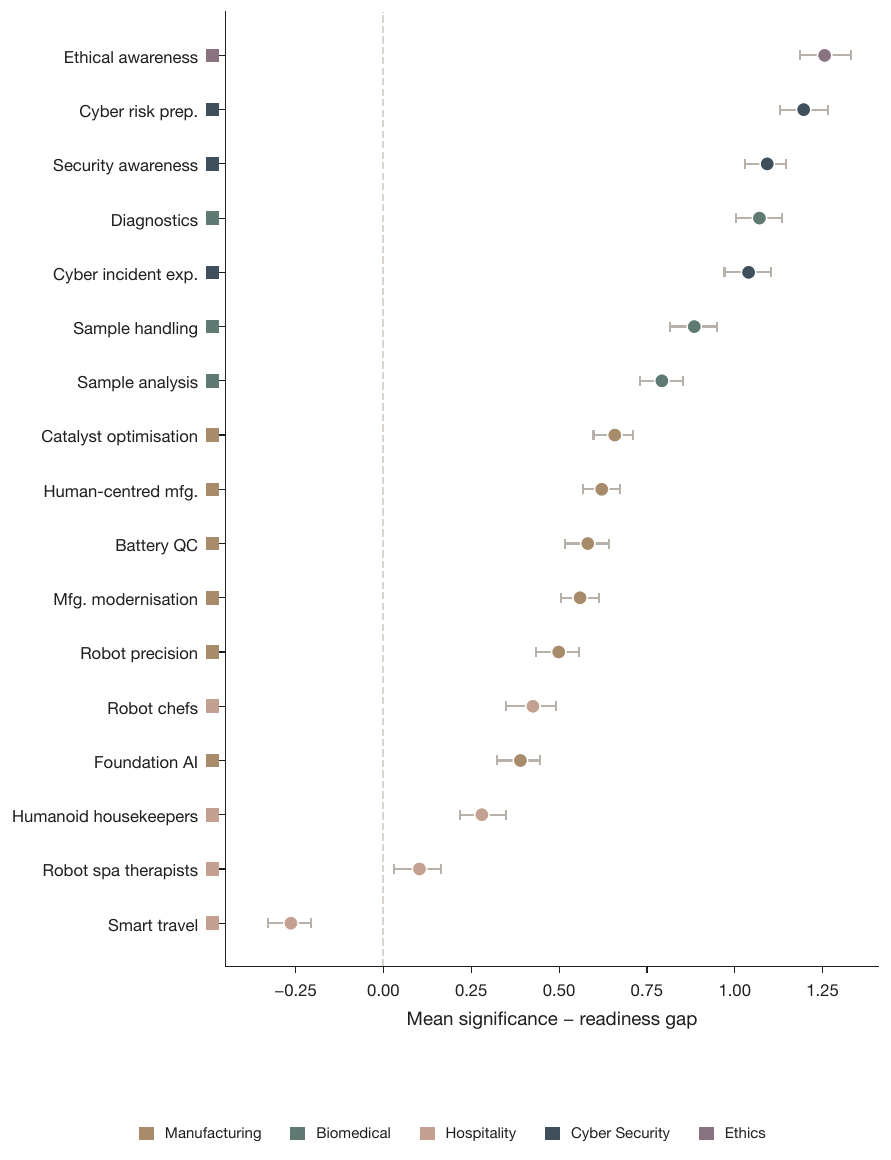}
\caption{Ranked significance--readiness gaps}
\label{fig:gapswithin_a}
\end{subfigure}\hfill
\begin{subfigure}[t]{0.40\textwidth}
\centering
\includegraphics[width=\linewidth]{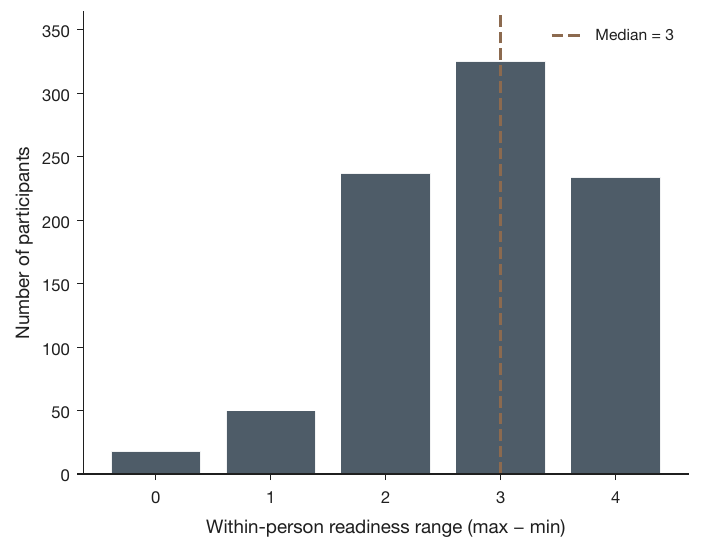}
\caption{Within-person readiness ranges}
\label{fig:gapswithin_b}
\end{subfigure}
\caption{Descriptive gaps and within-person variation. (a)~Ranked mean significance-readiness gaps with participant-bootstrap 95\% confidence intervals. (b)~Distribution of within-person readiness ranges among complete-deck participants.}
\label{fig:gapswithin}
\end{figure*}

Figure~\ref{fig:gapswithin}a presents the significance-readiness gaps with participant-bootstrap 95\% confidence intervals. The ordering should be interpreted descriptively rather than as a strict ranking, particularly where confidence intervals overlap. The confidence intervals nevertheless show that the size and uncertainty of the gap vary across challenges, reinforcing that the relationship between significance and readiness is not uniform across the challenge set.

Mean ratings also do not reveal how consistently respondents judge the same challenge. Card-level standard deviations (SDs), interquartile ranges (IQRs) and full readiness Likert distributions (Supplementary Table~S12; Supplementary Fig.~S3) show substantial differences in the spread of responses across cards. Challenges with similar mean readiness can therefore represent very different patterns of judgement. A moderate mean may reflect broad agreement that preparedness is moderate, or it may result from substantially different views that happen to average to the same value. Challenge means should therefore be interpreted together with the distribution of individual responses.

Challenge specificity becomes even clearer when the same respondents are compared across cards. Among the 864 participants who completed all 17 challenges, mean participant-level readiness is $3.15$, with a between-person SD of $0.70$. More importantly, the median difference between each respondent's highest and lowest readiness ratings is three points on the five-point scale, with a mean range of $2.82$. Only 7.9\% of respondents vary by one point or less across the full challenge set, whereas 36.6\% have a within-person readiness SD of at least one point. Respondents therefore do not simply hold a stable view that AI and robotics are generally `ready' or `not ready'; their assessments change substantially depending on the challenge being considered.

\begin{table*}[tbp]
\centering
\caption{Core within-between model of perceived readiness (participant-clustered robust standard errors; challenge-family and participant controls included).}
\label{tab:core}
\scriptsize
\input{scr_tables_core.tex}
\end{table*}

Together with the ICC, these results provide empirical support for distinguishing stable differences between respondents from challenge-specific variation within the same respondent. This distinction is central to the subsequent analysis because relationships observed across challenges within the same person may not be visible when respondents are compared only through their average ratings.

Exploratory Spearman correlations provide a consistent descriptive picture. The significance-readiness association is near zero for Ethical Awareness and weak for Cyber Security, but stronger in Biomedical, Manufacturing and Hospitality (Supplementary Table~S2). This further indicates that the relationship between perceived importance and preparedness depends on the challenge context rather than following a single pattern across AI and robotics.

\subsection{Challenge-specific complexity is associated with lower readiness}

Table~\ref{tab:core} reports the primary within-between model. Within-person significance is positively associated with readiness ($\hat\beta_{SW}=0.342$, $p<0.001$). More central to the preparedness argument, within-person complexity is negatively associated with readiness ($\hat\beta_{CW}=-0.205$, $p<0.001$): conditional on the other model terms, a card rated one point more complex than the respondent's own average is associated with approximately 0.21 points lower readiness.

\begin{table*}[tbp]
\centering
\caption{Primary robustness comparison for core within-person associations with perceived readiness.}
\label{tab:robust}
\scriptsize
\setlength{\tabcolsep}{2pt}
\renewcommand{\arraystretch}{0.92}

\input{scr_tables_robustness_main_v9.tex}

\renewcommand{\arraystretch}{1.0}
\setlength{\tabcolsep}{6pt}
\end{table*}

\begin{figure*}[btp]
\centering
\begin{subfigure}[t]{0.48\textwidth}
\centering
\includegraphics[width=\linewidth]{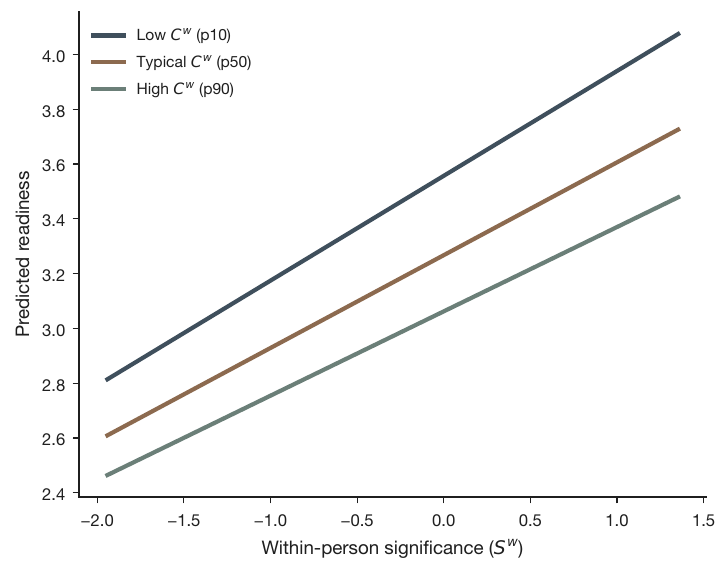}
\caption{Model-based interaction}
\label{fig:mechanism_a}
\end{subfigure}\hfill
\begin{subfigure}[t]{0.48\textwidth}
\centering
\includegraphics[width=\linewidth]{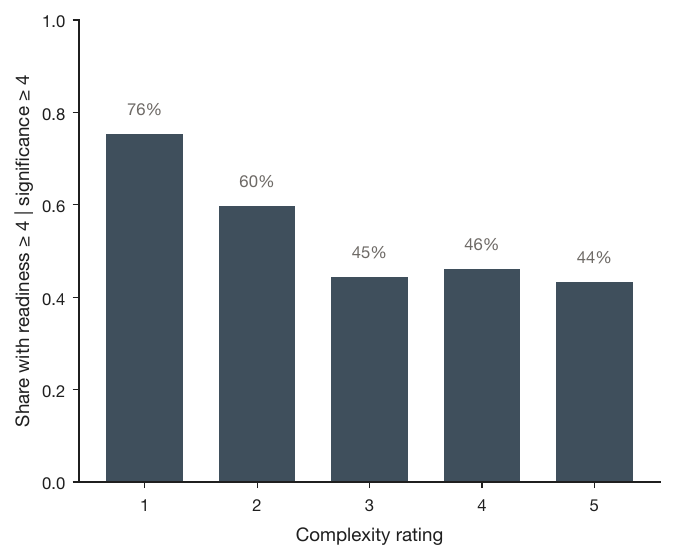}
\caption{High-significance readiness share}
\label{fig:mechanism_b}
\end{subfigure}
\caption{Complexity and the significance-readiness association. (a)~Model-based marginal association between within-person significance and readiness at low, typical and high within-person complexity. (b)~Descriptive share of high-significance responses ($S\geq4$) with readiness $\geq4$, shown separately for each complexity rating from 1 to 5.}
\label{fig:mechanism}
\end{figure*}

The between-person complexity coefficient, which captures the association between a respondent's average complexity rating and readiness, is smaller and not statistically significant ($\hat\beta_{CB}=-0.052$, $p=0.29$). This suggests that the negative association between complexity and readiness is mainly driven by how a particular challenge differs from a respondent's own usual rating, rather than by some respondents generally rating all challenges as more complex.

The interaction between within-person significance and complexity is also negative ($\hat\beta_{SC}=-0.031$, $p=0.004$). Although the effect is modest, it indicates that the positive association between significance and readiness becomes weaker when the same challenge is perceived as unusually complex by that respondent.

\begin{figure*}[tbp]
\centering
\begin{subfigure}[t]{0.48\textwidth}
\centering
\includegraphics[width=\linewidth]{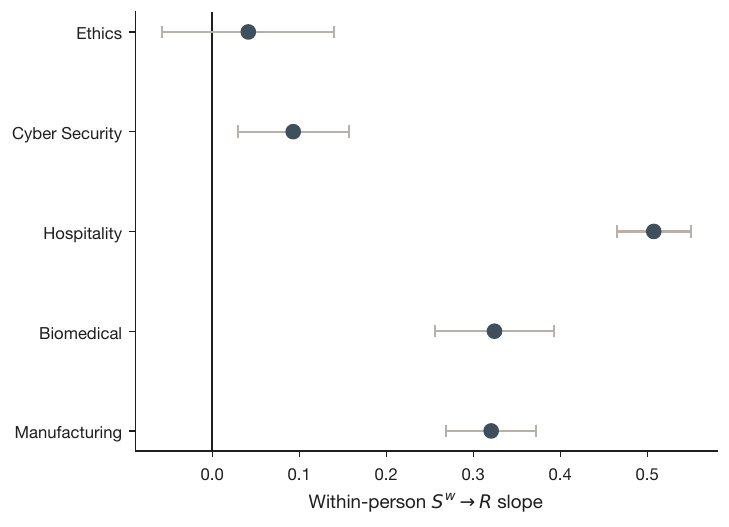}
\caption{Significance--readiness slopes}
\label{fig:familyslopes_a}
\end{subfigure}\hfill
\begin{subfigure}[t]{0.48\textwidth}
\centering
\includegraphics[width=\linewidth]{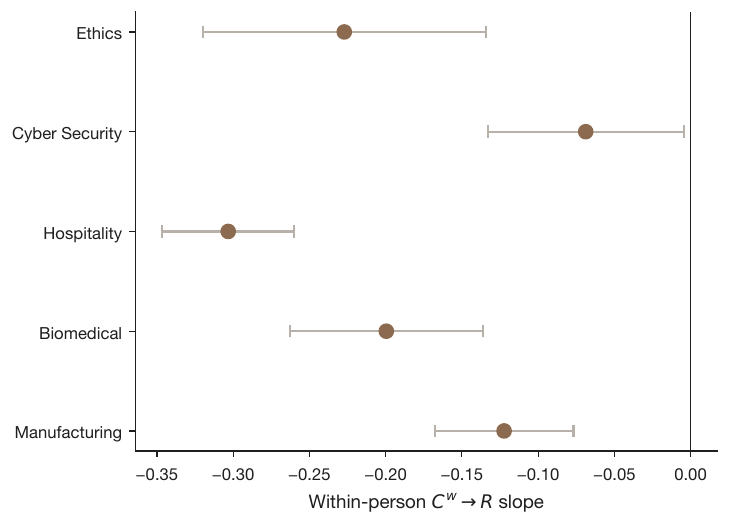}
\caption{Complexity--readiness slopes}
\label{fig:familyslopes_b}
\end{subfigure}
\caption{Challenge-family heterogeneity in within-person associations. (a)~Significance-readiness slopes. (b)~Complexity-readiness slopes.}
\label{fig:familyslopes}
\end{figure*}

Table~\ref{tab:robust} summarises the three primary robustness checks. The complete-deck within-between re-estimate is essentially identical to the main model ($\hat\beta_{CW}=-0.205$), indicating that incomplete person-means are not driving the result. Participant-and-card fixed effects attenuate the within-person complexity coefficient to approximately $-0.17$ but preserve its sign and significance after absorbing all stable respondent effects and stable card/position-specific intercept differences. Ordered logit and ordinal GEE likewise recover a negative complexity association (OR$\approx0.70$) and a positive significance association (OR$\approx1.8$-$1.9$), with a modestly negative within-person interaction. Because these checks agree with the primary model in direction and substantive interpretation, we retain the within-between linear coefficients as the main effect-size presentation. Extended pooled OLS (Supplementary Table~S5), ordered-logit detail (Supplementary Table~S9), crossed LMM and exploratory crossed cumulative-link mixed-model comparisons (Supplementary Methods~S-M) are reported only in the Supplementary Material; the pooled raw-$S$/$C$ benchmark should not be confused with earlier OLS specifications that omitted the interaction and/or used a different control set.


Figure~\ref{fig:mechanism}a presents marginal predictions from Eq.~\eqref{eq:core}. Figure~\ref{fig:mechanism}b provides a complementary empirical benchmark across all five complexity ratings: among responses with significance ratings of at least 4, the share with readiness $\geq4$ declines from approximately 76\% at complexity~1 and 60\% at complexity~2 to about 45\% at complexity~3 and remains near 43-46\% at complexities~4-5. The panel is descriptive and should not be interpreted as the causal effect of reducing complexity. Descriptively, the largest decline occurs between the lower and middle portions of the complexity scale, while the high-complexity categories show broadly similar proportions of high readiness.

Omnibus Wald tests show that the within-person relationships differ across challenge families. Equality of the complexity-readiness slopes is rejected ($\chi^{2}(4)=62.59$, $p<0.001$), as is equality of the significance-readiness slopes ($\chi^{2}(4)=143.10$, $p<0.001$). The $\chi^{2}$ values indicate the strength of evidence for heterogeneity rather than effect size. Family-specific estimates show stronger negative complexity-readiness associations in Hospitality (approximately $-0.30$), Ethical Awareness ($-0.23$) and Biomedical ($-0.20$), with a much smaller association in Cyber Security ($-0.07$) (Fig.~\ref{fig:familyslopes}b). Significance-readiness slopes also vary across families (Fig.~\ref{fig:familyslopes}a). Because Ethical Awareness contains only one card, its estimate is scenario-specific. Exact coefficients and confidence intervals are reported in Supplementary Table~S4.

\begin{figure*}[tbp]
\centering
\begin{subfigure}[t]{0.48\textwidth}
\centering
\includegraphics[width=\linewidth]{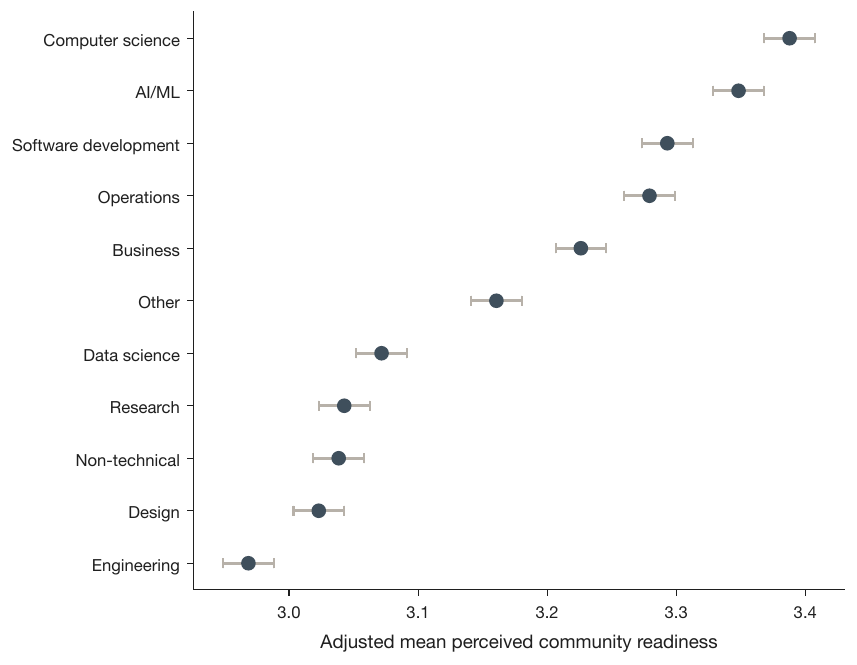}
\caption{Adjusted marginal readiness}
\label{fig:stakeholders_a}
\end{subfigure}\hfill
\begin{subfigure}[t]{0.48\textwidth}
\centering
\includegraphics[width=\linewidth]{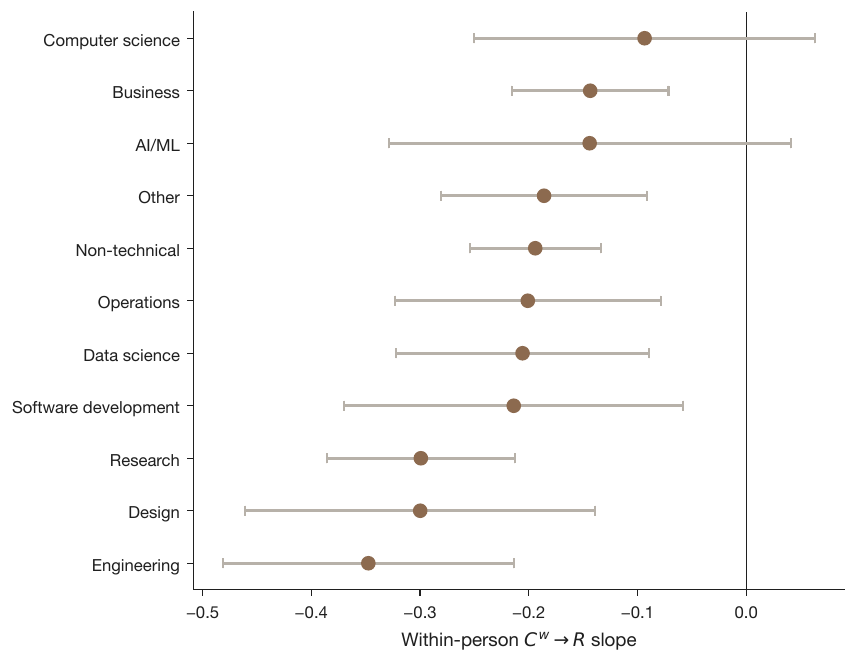}
\caption{Complexity--readiness slopes}
\label{fig:stakeholders_b}
\end{subfigure}
\caption{Stakeholder patterns. (a)~Adjusted marginal mean perceived community readiness by professional background, with 95\% confidence intervals (model-adjusted marginal means from the primary within-between model). (b)~Within-person complexity-readiness slopes by professional background (descriptive; omnibus interaction $p=0.11$).}
\label{fig:stakeholders}
\end{figure*}

\subsection{Preparedness levels and complexity sensitivity differ across stakeholders}

Stakeholder heterogeneity is relevant in two distinct ways: groups may differ in their average preparedness assessments, and they may differ in how strongly challenge-specific complexity is associated with readiness. Unadjusted professional-background, age and industry profiles are reported in Supplementary Table~S3.

An omnibus Wald test indicates that adjusted readiness differs across professional backgrounds after accounting for significance, complexity, challenge family and the other primary covariates ($\chi^{2}(10)=34.01$, $p<0.001$). Figure~\ref{fig:stakeholders}a therefore reports adjusted marginal mean readiness for every background rather than framing the demographic story as contrasts against a single reference cell. Adjusted marginal mean estimates are comparatively lower for engineering, non-technical, design, research and data-science backgrounds and higher for several computer-science/AI-oriented backgrounds, but the figure should be read as a full profile with uncertainty rather than as a set of multiplicity-adjusted pairwise contrasts. Treatment-coded contrasts versus the non-technical reference category ($n=288$; Supplementary Table~S7) and complementary age contrasts remain in the Supplementary Material; within-person complexity slopes by background are reported in Supplementary Table~S8. Omnibus tests do not support a strong joint age or education effect after the primary controls, while gender shows a modest omnibus association and is treated as exploratory.

The second stakeholder question concerns complexity sensitivity (Fig.~\ref{fig:stakeholders}b). Point estimates of within-person complexity penalties are largest among engineering (approximately $-0.35$), design ($-0.30$) and research ($-0.30$) respondents, and smallest for computer science (approximately $-0.09$). An omnibus test of the complexity$\times$background interactions does not reject equality of slopes at conventional levels ($\chi^{2}(10)=15.67$, $p=0.11$), so these subgroup differences are reported as descriptive heterogeneity rather than as confirmed differential slopes. Professional-background moderation of complexity sensitivity is therefore interpreted as a hypothesis-generating pattern, not as a supported interaction finding.

\subsection{Confidence and trust add diagnostic information on applied cards}

\begin{figure}[ptb]
\centering
\includegraphics[width=0.5\textwidth]{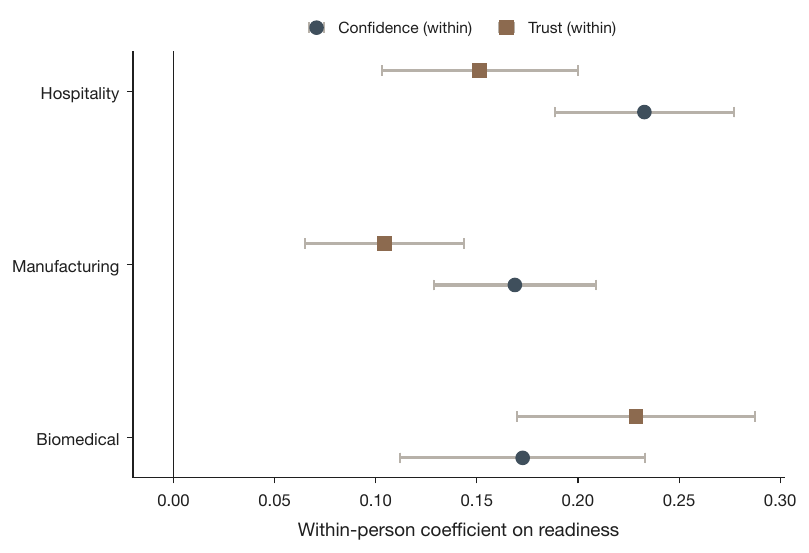}
\caption{Within-person conditional associations of confidence and trust with perceived readiness on applied challenge cards.}
\label{fig:diagnostics}
\end{figure}

On 11,469 applied-card responses, adding within- and between-person confidence, trust, familiarity, convenience and satisfaction increases in-sample $R^2$ from $0.208$ to $0.434$ and adjusted $R^2$ from $0.207$ to $0.433$ (partial $R^{2}$ of the perception block $=0.286$). Critically, the gain survives participant-level five-fold cross-validation: CV $R^{2}$ rises from $0.192$ to $0.419$ (increases of approximately $0.23$ in both in-sample and held-out $R^{2}$), showing that the improvement extends to held-out participants rather than being confined to in-sample fit (Table~\ref{tab:modelb}). Cross-validation strengthens the predictive argument but does not eliminate common-method correlation among confidence, trust, satisfaction and readiness; nor does participant-level CV test transfer to unseen challenge cards. Within-person confidence (approximately $0.20$) and trust ($0.15$) retain positive conditional associations in the pooled model. Point estimates suggest a relatively stronger trust association in Biomedical and stronger confidence associations in Manufacturing and Hospitality (Fig.~\ref{fig:diagnostics}; Supplementary Table~S10c). Formal tests of $H_0:\beta_{\mathrm{trust}}-\beta_{\mathrm{confidence}}=0$ within applied-card family do not distinguish the Biomedical coefficients ($p=0.31$), whereas Manufacturing ($p=0.039$) and Hospitality ($p=0.048$) both show larger confidence than trust coefficients at the conventional 5\% threshold; because these are targeted secondary comparisons, the results are interpreted cautiously (Supplementary Table~S10c). Because these perceptual items are correlated and measured contemporaneously with readiness, the comparison remains diagnostic rather than causal.

\begin{table}[tbp]
\centering
\caption{Incremental model performance on the applied cards.}
\label{tab:modelb}
\scriptsize
\setlength{\tabcolsep}{2pt}
\renewcommand{\arraystretch}{0.90}

\input{scr_tables_modelb_v8.tex}

\renewcommand{\arraystretch}{1.0}
\setlength{\tabcolsep}{6pt}
\end{table}

Exploratory free-text analysis provides complementary context. Among optional comments meeting predefined configuration thresholds ($S\geq4$ with $R\leq2$ versus $S\geq4$ with $R\geq4$; $n=198$ and $n=316$ usable comments), automated lexicon/dictionary scoring indicates that high-significance/low-readiness comments are more negative and more frequently tagged with privacy/security concerns than high-significance/high-readiness comments, whereas the latter more often emphasise efficiency and benefits. Human-experience themes appear in both groups. These results are retained as supporting evidence rather than a separate contribution because free-text participation is optional and therefore selected (Supplementary Methods~S-FT; Supplementary Fig.~S2 and Table~S11).

\section{Discussion}

\begin{table*}[tbp]
\centering
\caption{Policy diagnostic framework based on significance-complexity-readiness configurations and stakeholder evidence.}
\label{tab:policy}
\scriptsize
\rowcolors{2}{white}{rowblue}
\begin{tabularx}{\textwidth}{p{2.9cm}p{3.2cm}X p{3.3cm}}
\toprule
Observed pattern & Interpretation & What to investigate & Candidate response to test \\
\midrule
High $S$, high $C$, low $R$ & Important, difficult and comparatively lower in perceived preparedness & Capability, resources, coordination, assurance & Capacity building, staged implementation, assurance support \\
High $S$, low $C$, low $R$ & Preparedness shortfall not readily explained by complexity & Skills, access, institutional resources, ownership & Training, resource allocation, coordination \\
Strong negative $C^{w}\!\rightarrow\!R$ & Challenge-specific complexity sensitivity & Possible challenge-specific implementation burden & Simplification, co-design, testbeds, staged rollout \\
Low confidence with low $R$ & Possible self-efficacy/capability-related constraint & Practical experience, perceived capability, access to training and integration support & Hands-on training, demonstrations, test environments \\
Low trust with low $R$ & Possible assurance-related constraint & Validation, accountability, transparency, oversight & Evidence, governance, audit and escalation mechanisms \\
Persistent stakeholder contrast & Unequal preparedness perceptions & Exposure, role, resources and institutional context & Targeted engagement and support rather than uniform messaging \\
Low $S$, high $C$, low $R$ & Difficult challenge with lower perceived priority & Strategic value and use-case fit & Reconsider resource commitment or deployment case \\
\bottomrule
\end{tabularx}
\rowcolors{2}{white}{white}
\end{table*}

The results support a view of AI and robotics preparedness as challenge-specific rather than as a single stable readiness score. The same respondents vary substantially across the 17 challenge cards, and the joint SCR landscape separates challenges that are consequential and comparatively prepared from those that are consequential, difficult and comparatively lower in perceived preparedness. This extends person-level technology-readiness perspectives \citep{Parasuraman2000} by showing why a general disposition alone is insufficient for challenge-level forecasting, and it complements technology-acceptance models \citep{Davis1989,Venkatesh2003,Venkatesh2012} by focusing on perceived preparedness rather than behavioural intention.

The most consistent result concerns complexity. Once each respondent's average complexity level across cards is separated from card-specific deviations around that average, a challenge judged unusually complex for that same respondent is also judged less ready. By contrast, respondents whose average complexity rating is higher across the card set are not systematically more pessimistic about readiness. Interpreted carefully, significance is relevant to prioritisation and complexity is an indicator of potential implementation burden, but those interpretations go beyond the literal survey wording and should not be confused with direct measures of policy salience or organisational burden. The negative significance-complexity interaction is comparatively small but points in the same direction: consequentiality and preparedness are less closely aligned for challenges that a respondent sees as unusually difficult. The stability of the complexity estimate in the complete-deck sample further indicates that the result is not driven by participants' person means being calculated from different subsets of cards.

The challenge-family interactions show that these relationships are not uniform across AI and robotics contexts. Both the complexity-readiness and significance-readiness slopes differ jointly across challenge families ($p<0.001$ for both omnibus interaction tests). This reinforces the central forecasting argument: perceived preparedness should not be inferred from generic attitudes towards AI, because the relationship between perceived difficulty, consequentiality and preparedness depends materially on the type of challenge under consideration.

Stakeholder heterogeneity adds a second layer. Adjusted marginal readiness differs across professional backgrounds. Descriptive complexity slopes are steeper among engineering, design and research respondents, but the omnibus complexity$\times$background test is not decisive ($p=0.11$), so differences in complexity sensitivity remain suggestive rather than conclusive. One plausible interpretation is that some professional groups are more exposed to integration, validation, workflow or resource dependencies, but this mechanism is not directly measured. The policy value of the demographic analysis is therefore not to label groups as ready or unready, but to identify where preparedness assessments diverge enough to warrant targeted investigation.

Confidence and trust provide a third diagnostic layer on the applied cards. Their substantial in-sample and cross-validated contribution suggests that two challenges with similar SCR profiles may still differ in the perceptual conditions surrounding implementation. Point estimates lean toward a relatively stronger trust association in Biomedical and stronger confidence associations in Manufacturing and Hospitality-consistent with assurance- versus self-efficacy/capability-related hypotheses-but only the Manufacturing and Hospitality coefficient differences are statistically distinguished (Supplementary Table~S10c), and the cross-sectional design cannot establish that changing either perception would change preparedness.

For technological forecasting and policy, the practical implication is to evaluate significance, complexity and readiness jointly and then interpret the configuration in stakeholder context. Table~\ref{tab:policy} translates the observed patterns into questions to investigate and candidate responses to test. The table is deliberately diagnostic: it does not claim that a particular statistical pattern proves a specific intervention will work.

Several limitations bound these interpretations. The measures are self-reported and cross-sectional, and Readiness represents perceived community preparedness rather than independently audited organisational capability. Significance, Complexity and Readiness were collected using the same survey interface and may therefore share common-method variance. Cards were presented in a fixed order, meaning that card identity cannot be separated fully from presentation-order effects. The Prolific sample is not a probability sample of the UK workforce, and Ethical Awareness is represented by a single card. Some demographic groups are small and experience contains substantial missingness. Confidence, Trust and Readiness were measured contemporaneously, so their associations should not be interpreted causally. The complete-deck, participant and card fixed-effects, ordered-logit and ordinal-GEE robustness checks support the direction of the main within-person findings, but the study identifies diagnostic associations rather than the causal effects of training, assurance, resource provision or reductions in Complexity.

\section{Conclusion}

AI and robotics Readiness should not be treated as a single general score. The central finding is that a challenge perceived as unusually complex by the same respondent is associated with lower Readiness, while the corresponding relationship is not evident when respondents are compared using their average Complexity ratings. Aggregating across people can therefore conceal challenge-specific barriers that matter for adoption.

For policymakers and organisations, this changes what a Readiness assessment should be used for. Rather than asking only whether a sector, workforce or community is `ready for AI', decision makers should ask: \emph{ready for which challenge, according to which groups, and what appears to be constraining Readiness?} Significance, Complexity and Readiness should be considered jointly, while professional context, Confidence and Trust can provide additional clues about whether the next investigation should focus on implementation difficulty, capability, resources or assurance.

The practical value of the framework is therefore diagnostic. It can help policymakers target adoption support, literacy programmes and assurance activity more precisely, and help organisations identify where different professional groups perceive barriers that may be hidden by overall averages. The evidence does not prescribe a particular intervention, but it provides a more informative basis for deciding where support should be tested and why.

\section*{Declaration of competing interest}
The authors declare that they have no known competing financial interests or personal relationships that could have appeared to influence the work reported in this paper.

\section*{Data availability}
Additional tables, figures and methods notes are provided in the Supplementary Information appended to this document. Repository and data-access details will be inserted following the study's disclosure and ethics requirements.

\bibliography{references,references_additions_v2}

\clearpage
\let\table\tableorg
\let\endtable\endtableorg
\begin{appendices}

\renewcommand{\thefigure}{S\arabic{figure}}
\renewcommand{\thetable}{S\arabic{table}}
\renewcommand{\thesection}{S\arabic{section}}
\setcounter{figure}{0}
\setcounter{table}{0}
\setcounter{section}{0}

\clearpage
\input{supplement_freetext_methods.tex}

\clearpage
\onecolumn

\section{Participant characteristics}
\subsection*{Table S1. Full participant overview}

\centering
\input{scr_supplement/S1_sample_overview.tex}

\clearpage
\twocolumn
\subsection*{Table S1b. Education and industry detail}
\fitab{\input{scr_supplement/S1b_education_industry.tex}}

\section{Exploratory associations}
\subsection*{Table S2. Pairwise Spearman SCR associations by challenge family}
\fitab{\input{scr_supplement/S2_spearman.tex}}

\clearpage
\section{Unadjusted demographic profiles}
\subsection*{Table S3. Unadjusted significance, complexity, readiness and gap by demographic category}
{\scriptsize
\fitab{\input{scr_supplement/S3_unadjusted_demographics.tex}}
}

\clearpage
\section{Challenge-family within-person slopes}
\subsection*{Table S4. Challenge-family within-person slopes for significance and complexity}
\fitab{\input{scr_supplement/S4_application_slopes.tex}}

\clearpage
\section{Pooled OLS benchmark}
\subsection*{Table S5. Original pooled OLS readiness model (selected/full coefficient listing)}
{\scriptsize
\fitab{\input{scr_supplement/S5_pooled_OLS.tex}}
}

\subsection*{Table S5b. Pooled OLS fit summary}
\input{scr_supplement/S5_pooled_OLS_meta.tex}

\clearpage
\section{Variance decomposition and challenge-family residuals}
\subsection*{Table S6. Intraclass correlation for readiness}
\input{scr_supplement/S6_icc.tex}

\subsection*{Table S6b. Challenge-family residual/means summary}
\input{scr_supplement/S6_application_residuals.tex}

\begin{figure}[!htbp]
\centering
\includegraphics[width=0.85\textwidth]{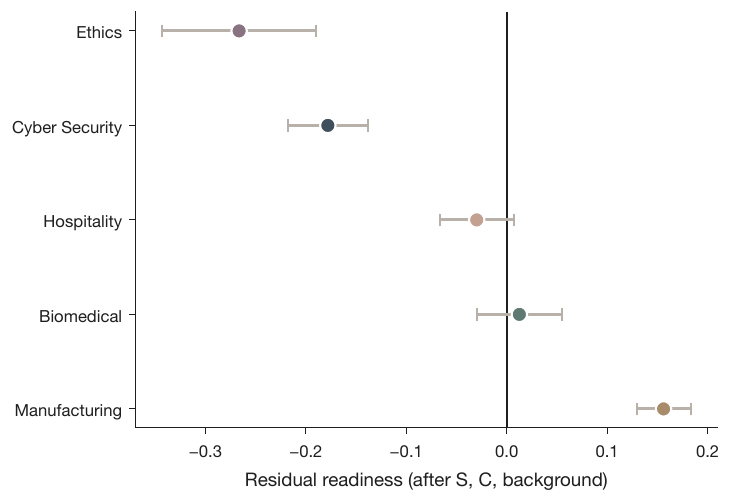}
\caption{Challenge-family residual or random-effect summary for readiness (Supplementary Fig.~S1).}
\label{fig:S1}
\end{figure}

\clearpage
\section{Full adjusted coefficients}
\subsection*{Table S7. Full coefficient listing from the core within--between readiness model}
{\scriptsize
\fitab{\input{scr_supplement/S7_full_adjusted_coefs.tex}}
}

\clearpage
\section{Professional-background complexity sensitivity}
\subsection*{Table S8. Within-person complexity--readiness slopes by professional background}
\fitab{\input{scr_supplement/S8_background_complexity_slopes.tex}}

\section{Ordered-logit robustness}
\subsection*{Table S9. Cumulative ordered-logit core predictors}
\fitab{\input{scr_supplement/S9_ordered_logit.tex}}

\clearpage
\section{Applied-card perceptual models}
\subsection*{Table S10. Model B comparison by applied subset}
\fitab{\input{scr_supplement/S10_modelB_comparison.tex}}

\subsection*{Table S10b. Selected Model B perception coefficients}
{\scriptsize
\fitab{\input{scr_supplement/S10b_modelB_coefs.tex}}
}

\clearpage
\subsection*{Table S10c. Within-family contrast of trust vs.\ confidence coefficients}
\noindent\small Wald tests of $H_0:\beta_{\mathrm{trust}^{w}}-\beta_{\mathrm{confidence}^{w}}=0$ from sector-specific applied-card Model~B regressions for the three families that share the common confidence/trust battery (Biomedical, Manufacturing and Hospitality), with participant-clustered standard errors (same covariates as the main Model~B specification, excluding family indicators within each single-family subset). Cyber Security and Ethical Awareness cards are excluded because they use specialist follow-up items without equivalent trust items. Positive $\Delta$ means the trust coefficient is larger than the confidence coefficient.
\fitab{\input{scr_supplement/S10c_trust_conf_contrast.tex}}

\section{Free-text coding}
\subsection*{Table S11. Sentiment summaries by preparedness configuration}
\input{scr_supplement/S11_freetext_sentiment.tex}

\subsection*{Table S11b. Theme prevalence by preparedness configuration}
{\scriptsize
\fitab{\input{scr_supplement/S11_freetext_themes.tex}}
}

\clearpage
\begin{figure}[!htbp]
\centering
\includegraphics[width=0.9\textwidth]{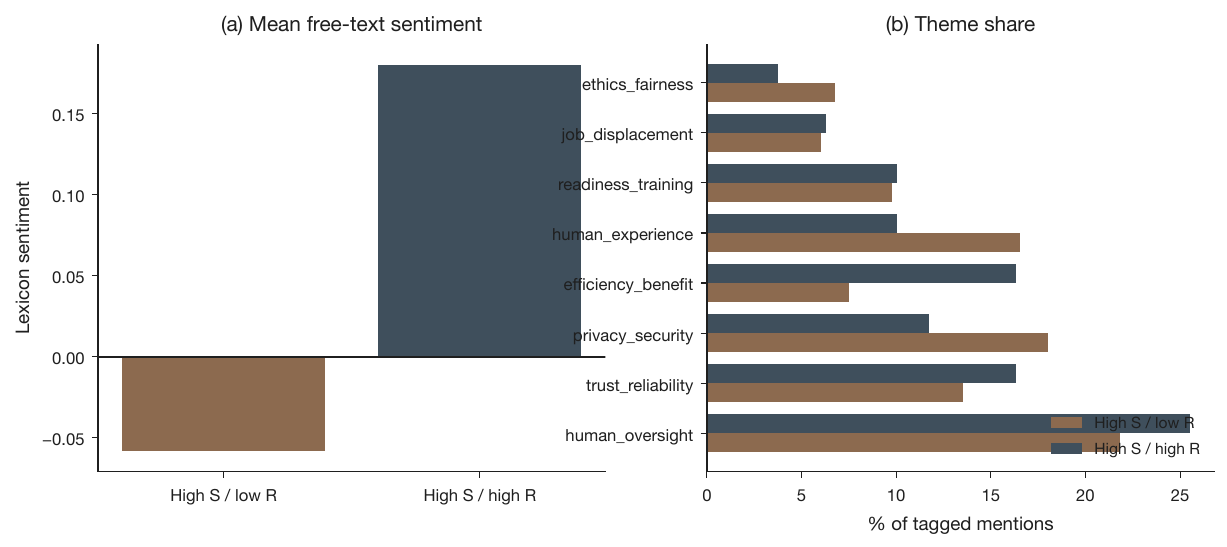}
\caption{Theme prevalence by high-significance/low-readiness versus high-significance/high-readiness free-text responses (Supplementary Fig.~S2).}
\label{fig:S2}
\end{figure}

\clearpage
\section{Same-card response dispersion (Table S12 and Fig.~S3)}
\label{sec:S12}
Card means alone do not show whether respondents agree about a given challenge.
Table~S12 reports the Mean, SD and IQR of significance, complexity and readiness for each of the 17 cards; Supplementary Fig.~S3 shows the full readiness Likert distribution for each card.
These summaries demonstrate that different people do not necessarily rate the same card identically, while leaving a fuller stakeholder$\times$challenge disagreement analysis outside the scope of the main manuscript.

\begin{table}[!htbp]
\centering
\caption{Card-level Mean, SD and IQR for significance, complexity and readiness (Supplementary Table~S12).}
\label{tab:S12}
\scriptsize
\fitab{\input{scr_supplement/S12_card_dispersion.tex}}
\end{table}

\clearpage
\begin{figure}[!htbp]
\centering
\includegraphics[width=0.95\textwidth]{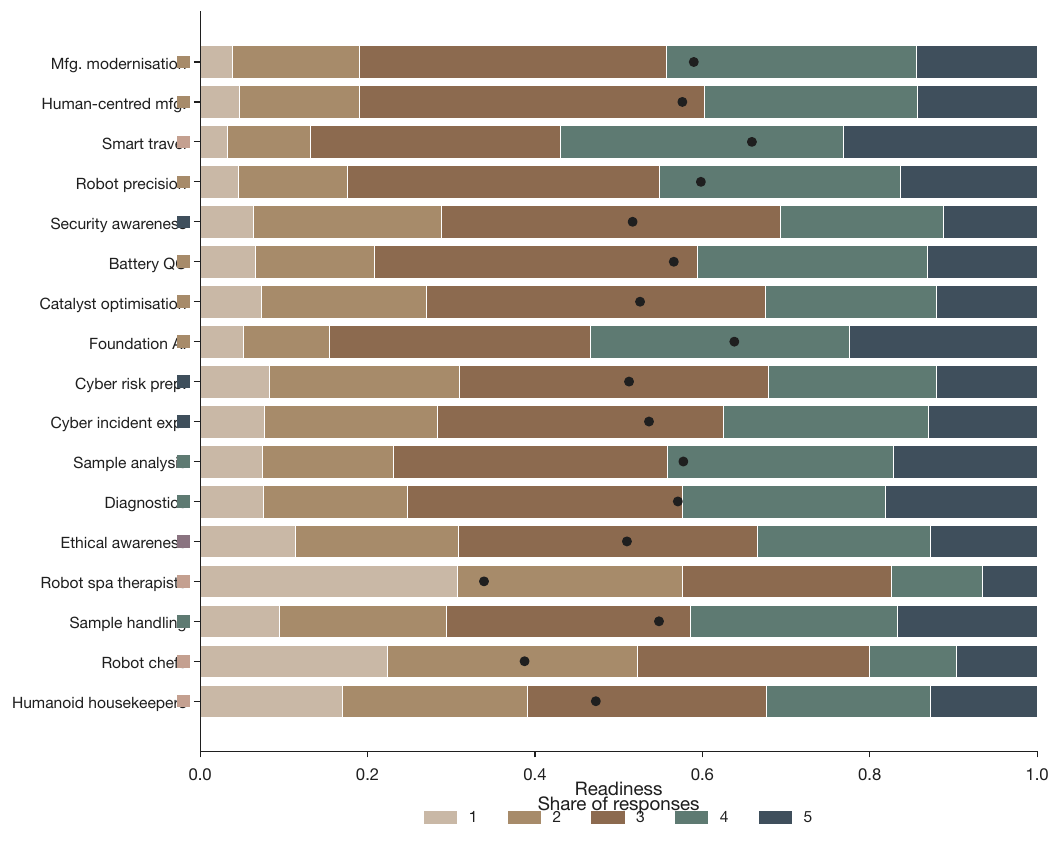}
\caption{Readiness rating distributions for each challenge card (Supplementary Fig.~S3).
Bars show the share of responses in Likert categories 1--5; black markers indicate the card mean mapped onto the unit interval.
Cards are ordered by readiness SD (lowest disagreement at the bottom).
Family colours mark the challenge family of each card.}
\label{fig:S3}
\end{figure}

\end{appendices}

\end{document}

%% file: scr_tables_card.tex
\rowcolors{2}{white}{rowblue}
\begin{tabular}{llrrrrr}
\toprule
Family & Challenge & $n$ & Sig. & Comp. & Ready & Gap \\
\midrule
Ethics & Ethical awareness & 888 & 4.30 & 3.94 & 3.04 & 1.26 \\
Cyber Security & Cyber risk preparedness & 949 & 4.25 & 4.03 & 3.05 & 1.20 \\
Cyber Security & Security awareness & 976 & 4.16 & 3.91 & 3.07 & 1.09 \\
Biomedical & Diagnostics & 870 & 4.35 & 4.21 & 3.28 & 1.07 \\
Cyber Security & Cyber incident experience & 918 & 4.18 & 3.90 & 3.14 & 1.04 \\
Biomedical & Sample handling & 867 & 4.08 & 3.90 & 3.19 & 0.88 \\
Biomedical & Sample analysis & 867 & 4.10 & 3.94 & 3.31 & 0.79 \\
Manufacturing & Catalyst optimisation & 881 & 3.76 & 3.83 & 3.10 & 0.66 \\
Manufacturing & Human-centred manufacturing & 874 & 3.93 & 3.78 & 3.30 & 0.62 \\
Manufacturing & Battery QC & 877 & 3.84 & 3.74 & 3.26 & 0.58 \\
Manufacturing & Manufacturing modernisation & 872 & 3.92 & 3.79 & 3.36 & 0.56 \\
Manufacturing & Robot precision & 868 & 3.89 & 3.77 & 3.39 & 0.50 \\
Hospitality & Robot chefs & 905 & 2.98 & 3.56 & 2.55 & 0.43 \\
Manufacturing & Foundation AI support & 872 & 3.94 & 3.54 & 3.55 & 0.39 \\
Hospitality & Humanoid housekeepers & 907 & 3.17 & 3.37 & 2.89 & 0.28 \\
Hospitality & Robot spa therapists & 898 & 2.46 & 3.27 & 2.36 & 0.10 \\
Hospitality & Smart travel & 911 & 3.37 & 2.98 & 3.64 & -0.26 \\
\bottomrule
\end{tabular}
\rowcolors{2}{white}{white}

%% file: scr_tables_core.tex
\rowcolors{2}{white}{rowblue}
\begin{tabular}{lrrr}
\toprule
Predictor & Coef. & 95\% CI & $p$ \\
\midrule
Significance - within & $0.342$ & $[0.313,0.371]$ & $<.001$ \\
Significance - between (person mean) & $0.491$ & $[0.400,0.581]$ & $<.001$ \\
Complexity - within & $-0.205$ & $[-0.235,-0.175]$ & $<.001$ \\
Complexity - between (person mean) & $-0.052$ & $[-0.149,0.044]$ & $0.290$ \\
Significance × Complexity - within & $-0.031$ & $[-0.052,-0.010]$ & $0.004$ \\
\bottomrule
\end{tabular}
\rowcolors{2}{white}{white}

%% file: scr_tables_robustness_main_v9.tex
\rowcolors{2}{white}{rowblue}

\begin{tabularx}{0.86\textwidth}{@{}Xccc@{}}
\toprule
Model & $S^{w}$ & $C^{w}$ & $S^{w}\times C^{w}$ \\
\midrule
Main linear
& $0.342$ $(<0.001)$
& $-0.205$ $(<0.001)$
& $-0.031$ $(0.004)$ \\

Complete deck
& $0.342$ $(<0.001)$
& $-0.205$ $(<0.001)$
& $-0.030$ $(<0.01)$ \\

Participant + card FE
& $0.296$ $(<0.001)$
& $-0.167$ $(<0.001)$
& $-0.028$ $(<0.01)$ \\

Ordered logit (OR)
& $1.86$ $(<0.001)$
& $0.70$ $(<0.001)$
& $0.93$ $(<0.001)$ \\

Ordinal GEE (OR)
& $1.80$ $(<0.001)$
& $0.70$ $(<0.001)$
& $0.96$ $(<0.01)$ \\
\bottomrule
\end{tabularx}

\vspace{1pt}

\begin{minipage}{0.86\textwidth}
\tiny
Values in parentheses are $p$-values. Linear models report coefficients; ordinal models report odds ratios. FE = fixed effects; GEE = generalised estimating equations.
\end{minipage}

\rowcolors{2}{white}{white}

%% file: scr_tables_modelb_v8.tex
\rowcolors{2}{white}{rowblue}

\begin{tabularx}{0.98\columnwidth}{@{}Xcc@{}}
\toprule
 & B0 & B1 \\
\midrule
$R^2$              & 0.208 & 0.434 \\
Adj.\ $R^2$        & 0.207 & 0.433 \\
CV $R^2$           & 0.192 & 0.419 \\
$\Delta R^2$       & --    & $\approx0.23$ \\
Partial $R^2$      & --    & 0.286 \\
Conf.\ / Trust     & --    & 0.199 / 0.155 \\
\bottomrule
\end{tabularx}

\vspace{1pt}

\begin{minipage}{0.98\columnwidth}
\tiny
B0 = base model; B1 = additional perceptions; CV = participant-level five-fold cross-validation. Conf./Trust are within-person coefficients.
\end{minipage}

\rowcolors{2}{white}{white}

%% file: supplement_freetext_methods.tex
\section*{Supplementary Methods S-FT: Exploratory free-text definitions}
\label{supp:freetext-methods}

Optional free-text comments were collected on applied cards only and are analysed exploratorily as complementary context, not as a primary inference target. Comments were retained if, after stripping whitespace, length was at least 10 characters and the text was not a filler token (e.g.\ ``n/a'', ``none'', ``no comment'', ``idk'', ``test'').

\paragraph{Configuration groups.}
Only high-significance comments enter the configuration comparison reported in the main text:
\begin{itemize}
\item \textbf{High significance / low readiness} ($S\geq4$ and $R\leq2$);
\item \textbf{High significance / high readiness} ($S\geq4$ and $R\geq4$).
\end{itemize}
Intermediate readiness values ($R=3$) and low-significance comments ($S\leq3$) are excluded from this two-group contrast. In the analytical sample, $n=198$ usable comments fall in the high-significance/low-readiness group and $n=316$ in the high-significance/high-readiness group.

\paragraph{Sentiment.}
Sentiment scores are dictionary-based rather than manually coded. Tokens are lower-cased alphabetic words. Positive and negative counts use the NLTK Opinion Lexicon when available, otherwise a small hardcoded fallback lexicon. The comment-level score is
\[
\frac{n_{\mathrm{pos}}-n_{\mathrm{neg}}}{n_{\mathrm{pos}}+n_{\mathrm{neg}}}
\]
when at least one lexicon hit is present, and $0$ otherwise. Group means are arithmetic means of comment-level scores. No multi-coder reliability coefficient is reported because coding is fully automated.

\paragraph{Themes.}
Theme tags are likewise dictionary-based: a predefined keyword list assigns zero or more theme labels to each comment (e.g.\ \textit{privacy\_security}, \textit{efficiency\_benefit}, \textit{human\_oversight}/\textit{human\_experience}, \textit{trust\_reliability}). Theme shares within each configuration group are percentages of tagged theme mentions, not of comments; a comment may contribute multiple themes. Theme dictionaries and cleaning rules are archived with the analysis code. Because participation is optional, free-text contrasts should be interpreted as selected supportive context rather than as population estimates of stakeholder opinion.

%% file: scr_supplement/S1_sample_overview.tex
\scriptsize
\setlength{\tabcolsep}{2.5pt}
\renewcommand{\arraystretch}{0.82}

\begin{minipage}[t]{0.30\textwidth}
\vspace{0pt}

\textbf{Overall}

\rowcolors{2}{white}{rowblue}
\begin{tabularx}{\linewidth}{@{}Xrr@{}}
\toprule
 & $n$ & \% \\
\midrule
Participants retained & 982 & 100.0 \\
Card responses & 15,200 & -- \\
Complete 17-card decks & 864 & -- \\
\bottomrule
\end{tabularx}
\rowcolors{2}{white}{white}

\vspace{4pt}
\textbf{Gender}

\rowcolors{2}{white}{rowblue}
\begin{tabularx}{\linewidth}{@{}Xrr@{}}
\toprule
 & $n$ & \% \\
\midrule
Female & 494 & 50.3 \\
Male & 477 & 48.6 \\
Non-binary & 8 & 0.8 \\
Prefer not to say & 3 & 0.3 \\
\bottomrule
\end{tabularx}
\rowcolors{2}{white}{white}

\vspace{4pt}
\textbf{Age}

\rowcolors{2}{white}{rowblue}
\begin{tabularx}{\linewidth}{@{}Xrr@{}}
\toprule
 & $n$ & \% \\
\midrule
18-24 & 134 & 13.6 \\
25-34 & 360 & 36.7 \\
35-44 & 272 & 27.7 \\
45-54 & 123 & 12.5 \\
55-64 & 76 & 7.7 \\
65+ & 17 & 1.7 \\
\bottomrule
\end{tabularx}
\rowcolors{2}{white}{white}

\vspace{4pt}
\textbf{Experience}

\rowcolors{2}{white}{rowblue}
\begin{tabularx}{\linewidth}{@{}Xrr@{}}
\toprule
 & $n$ & \% \\
\midrule
0-2 & 114 & 11.6 \\
3-5 & 155 & 15.8 \\
6-10 & 152 & 15.5 \\
11-15 & 90 & 9.2 \\
16-20 & 65 & 6.6 \\
20+ & 152 & 15.5 \\
Missing & 254 & 25.9 \\
\bottomrule
\end{tabularx}
\rowcolors{2}{white}{white}

\end{minipage}
\hfill
\begin{minipage}[t]{0.32\textwidth}
\vspace{0pt}

\textbf{Professional background}

\rowcolors{2}{white}{rowblue}
\begin{tabularx}{\linewidth}{@{}Xrr@{}}
\toprule
 & $n$ & \% \\
\midrule
Non-technical & 288 & 29.3 \\
Business & 163 & 16.6 \\
Other & 117 & 11.9 \\
Research & 76 & 7.7 \\
Operations & 73 & 7.4 \\
Engineering & 53 & 5.4 \\
Computer science & 49 & 5.0 \\
Data science & 48 & 4.9 \\
Software development & 46 & 4.7 \\
AI/ML & 38 & 3.9 \\
Design & 31 & 3.2 \\
\bottomrule
\end{tabularx}
\rowcolors{2}{white}{white}

\vspace{4pt}
\textbf{Education}

\rowcolors{2}{white}{rowblue}
\begin{tabularx}{\linewidth}{@{}Xrr@{}}
\toprule
 & $n$ & \% \\
\midrule
Bachelor's & 415 & 42.3 \\
Master's & 233 & 23.7 \\
Some college & 160 & 16.3 \\
High school & 91 & 9.3 \\
Associate & 31 & 3.2 \\
PhD & 29 & 3.0 \\
Professional & 19 & 1.9 \\
Other & 4 & 0.4 \\
\bottomrule
\end{tabularx}
\rowcolors{2}{white}{white}

\vspace{4pt}
\textbf{Application family responses}

\rowcolors{2}{white}{rowblue}
\begin{tabularx}{\linewidth}{@{}Xrr@{}}
\toprule
 & $n$ & \% \\
\midrule
Biomedical & 2,604 & 17.1 \\
Cyber Security & 2,843 & 18.7 \\
Ethics & 888 & 5.8 \\
Hospitality & 3,621 & 23.8 \\
Manufacturing & 5,244 & 34.5 \\
\bottomrule
\end{tabularx}
\rowcolors{2}{white}{white}

\end{minipage}
\hfill
\begin{minipage}[t]{0.32\textwidth}
\vspace{0pt}

\textbf{Industry}

\rowcolors{2}{white}{rowblue}
\begin{tabularx}{\linewidth}{@{}Xrr@{}}
\toprule
 & $n$ & \% \\
\midrule
Technology & 136 & 13.8 \\
Other & 130 & 13.2 \\
Education & 105 & 10.7 \\
Healthcare & 101 & 10.3 \\
Retail & 100 & 10.2 \\
Finance & 76 & 7.7 \\
Student & 67 & 6.8 \\
Government & 64 & 6.5 \\
Media & 41 & 4.2 \\
Manufacturing & 31 & 3.2 \\
Engineering & 29 & 3.0 \\
Research & 22 & 2.2 \\
Retired & 22 & 2.2 \\
Consulting & 20 & 2.0 \\
Legal & 17 & 1.7 \\
Tourism & 15 & 1.5 \\
Energy & 6 & 0.6 \\
\bottomrule
\end{tabularx}
\rowcolors{2}{white}{white}

\end{minipage}

\renewcommand{\arraystretch}{1.0}
\setlength{\tabcolsep}{6pt}

%% file: scr_supplement/S1b_education_industry.tex
\rowcolors{2}{white}{rowblue}
\begin{tabular}{llrr}
\toprule
Block & Category & $n$ & \% \\
\midrule
Education & bachelors & 415 & 42.3 \\
Education & masters & 233 & 23.7 \\
Education & some college & 160 & 16.3 \\
Education & high school & 91 & 9.3 \\
Education & associate & 31 & 3.2 \\
Education & phd & 29 & 3.0 \\
Education & professional & 19 & 1.9 \\
Education & other & 4 & 0.4 \\
Industry & technology & 136 & 13.8 \\
Industry & other & 130 & 13.2 \\
Industry & education & 105 & 10.7 \\
Industry & healthcare & 101 & 10.3 \\
Industry & retail & 100 & 10.2 \\
Industry & finance & 76 & 7.7 \\
Industry & student & 67 & 6.8 \\
Industry & government & 64 & 6.5 \\
Industry & media & 41 & 4.2 \\
Industry & manufacturing & 31 & 3.2 \\
Industry & engineering & 29 & 3.0 \\
Industry & research & 22 & 2.2 \\
Industry & retired & 22 & 2.2 \\
Industry & consulting & 20 & 2.0 \\
Industry & legal & 17 & 1.7 \\
Industry & tourism & 15 & 1.5 \\
Industry & energy & 6 & 0.6 \\
Experience & Missing & 254 & 25.9 \\
Experience & 3 5 & 155 & 15.8 \\
Experience & 20+ & 152 & 15.5 \\
Experience & 6 10 & 152 & 15.5 \\
Experience & 0 2 & 114 & 11.6 \\
Experience & 11 15 & 90 & 9.2 \\
Experience & 16 20 & 65 & 6.6 \\
\bottomrule
\end{tabular}
\rowcolors{2}{white}{white}

%% file: scr_supplement/S2_spearman.tex
\rowcolors{2}{white}{rowblue}
\begin{tabular}{lrrrrrrrrr}
\toprule
Family & $\rho_{S,R}$ & $p$ & $\rho_{C,R}$ & $p$ & $\rho_{C,G}$ & $p$ & $\rho_{S,C}$ & $p$ & $n$ \\
\midrule
Biomedical & $0.255$ & $<.001$ & $0.002$ & $0.924$ & $0.366$ & $<.001$ & $0.509$ & $<.001$ & 2604 \\
Cyber Security & $0.069$ & $<.001$ & $-0.010$ & $0.579$ & $0.257$ & $<.001$ & $0.394$ & $<.001$ & 2843 \\
Ethics & $-0.006$ & $0.859$ & $-0.111$ & $<.001$ & $0.361$ & $<.001$ & $0.457$ & $<.001$ & 888 \\
Hospitality & $0.420$ & $<.001$ & $-0.061$ & $<.001$ & $0.345$ & $<.001$ & $0.341$ & $<.001$ & 3621 \\
Manufacturing & $0.324$ & $<.001$ & $0.095$ & $<.001$ & $0.286$ & $<.001$ & $0.459$ & $<.001$ & 5244 \\
\bottomrule
\end{tabular}
\rowcolors{2}{white}{white}

%% file: scr_supplement/S3_unadjusted_demographics.tex
\rowcolors{2}{white}{rowblue}
\begin{tabular}{llrrrrr}
\toprule
Dimension & Category & $n$ & Sig. & Comp. & Ready & Gap \\
\midrule
Professional background & ai ml & 38 & 3.86 & 3.74 & 3.40 & 0.46 \\
Professional background & business & 163 & 3.82 & 3.79 & 3.23 & 0.59 \\
Professional background & computer science & 49 & 3.97 & 3.91 & 3.51 & 0.47 \\
Professional background & data science & 48 & 3.96 & 3.78 & 3.18 & 0.78 \\
Professional background & design & 31 & 3.74 & 3.86 & 2.96 & 0.78 \\
Professional background & engineering & 53 & 3.75 & 3.76 & 2.96 & 0.78 \\
Professional background & non technical & 288 & 3.72 & 3.67 & 2.98 & 0.74 \\
Professional background & operations & 73 & 3.86 & 3.74 & 3.31 & 0.54 \\
Professional background & other & 117 & 3.79 & 3.69 & 3.14 & 0.65 \\
Professional background & research & 76 & 3.86 & 3.70 & 3.07 & 0.79 \\
Professional background & software dev & 46 & 3.82 & 3.68 & 3.33 & 0.49 \\
Age & 18 24 & 134 & 3.74 & 3.62 & 3.21 & 0.53 \\
Age & 25 34 & 360 & 3.83 & 3.72 & 3.19 & 0.64 \\
Age & 35 44 & 272 & 3.81 & 3.76 & 3.12 & 0.69 \\
Age & 45 54 & 123 & 3.78 & 3.78 & 3.06 & 0.72 \\
Age & 55 64 & 76 & 3.84 & 3.86 & 3.02 & 0.82 \\
Age & 65+ & 17 & 3.66 & 3.70 & 3.04 & 0.62 \\
Education & associate & 31 & 3.69 & 3.62 & 3.12 & 0.57 \\
Education & bachelors & 415 & 3.83 & 3.75 & 3.17 & 0.66 \\
Education & high school & 91 & 3.66 & 3.68 & 3.06 & 0.60 \\
Education & masters & 233 & 3.87 & 3.76 & 3.21 & 0.66 \\
Education & other & 4 & 4.10 & 4.21 & 3.51 & 0.59 \\
Education & phd & 29 & 3.83 & 3.72 & 3.08 & 0.75 \\
Education & professional & 19 & 3.80 & 3.86 & 2.88 & 0.92 \\
Education & some college & 160 & 3.74 & 3.66 & 3.07 & 0.68 \\
Gender & female & 494 & 3.77 & 3.71 & 3.09 & 0.69 \\
Gender & male & 477 & 3.84 & 3.76 & 3.22 & 0.63 \\
Gender & non binary & 8 & 3.50 & 3.60 & 2.71 & 0.79 \\
Gender & prefer not to say & 3 & 3.90 & 4.31 & 2.67 & 1.24 \\
Industry & consulting & 20 & 3.76 & 3.90 & 3.09 & 0.67 \\
Industry & education & 105 & 3.76 & 3.72 & 3.04 & 0.72 \\
Industry & energy & 6 & 3.65 & 3.30 & 3.34 & 0.31 \\
Industry & engineering & 29 & 3.81 & 3.85 & 3.04 & 0.77 \\
Industry & finance & 76 & 3.78 & 3.69 & 3.24 & 0.54 \\
Industry & government & 64 & 3.89 & 3.78 & 3.11 & 0.78 \\
Industry & healthcare & 101 & 3.78 & 3.75 & 3.24 & 0.54 \\
Industry & legal & 17 & 3.86 & 3.64 & 3.10 & 0.76 \\
Industry & manufacturing & 31 & 3.91 & 4.01 & 3.33 & 0.58 \\
Industry & media & 41 & 3.60 & 3.68 & 2.91 & 0.69 \\
Industry & other & 130 & 3.77 & 3.66 & 3.01 & 0.76 \\
Industry & research & 22 & 3.90 & 3.60 & 3.09 & 0.81 \\
Industry & retail & 100 & 3.80 & 3.78 & 3.11 & 0.70 \\
Industry & retired & 22 & 3.88 & 3.81 & 3.29 & 0.59 \\
Industry & student & 67 & 3.75 & 3.57 & 3.16 & 0.59 \\
Industry & technology & 136 & 3.89 & 3.77 & 3.29 & 0.60 \\
Industry & tourism & 15 & 3.79 & 3.68 & 3.05 & 0.74 \\
Experience & 0 2 & 114 & 3.67 & 3.55 & 3.11 & 0.57 \\
Experience & 11 15 & 90 & 3.85 & 3.81 & 3.09 & 0.76 \\
Experience & 16 20 & 65 & 3.78 & 3.71 & 3.18 & 0.60 \\
Experience & 20+ & 152 & 3.75 & 3.77 & 3.00 & 0.75 \\
Experience & 3 5 & 155 & 3.89 & 3.76 & 3.28 & 0.61 \\
Experience & 6 10 & 152 & 3.87 & 3.77 & 3.25 & 0.63 \\
\bottomrule
\end{tabular}
\rowcolors{2}{white}{white}

%% file: scr_supplement/S4_application_slopes.tex
\rowcolors{2}{white}{rowblue}
\begin{tabular}{lrrrrrr}
\toprule
Family & $S^{w}\rightarrow R$ & 95\% CI & $p$ & $C^{w}\rightarrow R$ & 95\% CI & $p$ \\
\midrule
Manufacturing & $0.321$ & $[0.269,0.372]$ & $<.001$ & $-0.122$ & $[-0.168,-0.077]$ & $<.001$ \\
Biomedical & $0.324$ & $[0.256,0.392]$ & $<.001$ & $-0.200$ & $[-0.263,-0.136]$ & $<.001$ \\
Hospitality & $0.507$ & $[0.465,0.550]$ & $<.001$ & $-0.303$ & $[-0.347,-0.260]$ & $<.001$ \\
Cyber Security & $0.093$ & $[0.029,0.157]$ & $0.004$ & $-0.069$ & $[-0.133,-0.004]$ & $0.036$ \\
Ethics & $0.041$ & $[-0.058,0.140]$ & $0.411$ & $-0.227$ & $[-0.320,-0.134]$ & $<.001$ \\
\bottomrule
\end{tabular}
\rowcolors{2}{white}{white}

%% file: scr_supplement/S5_pooled_OLS.tex
\rowcolors{2}{white}{rowblue}
\begin{tabular}{lrrr}
\toprule
Term & Coef. & SE & $p$ \\
\midrule
Intercept & $1.861$ & $0.197$ & $<.001$ \\
C(sector, Treatment(reference='Manufacturing'))[T.Biomedical] & $-0.136$ & $0.027$ & $<.001$ \\
C(sector, Treatment(reference='Manufacturing'))[T.Cyber Security] & $-0.335$ & $0.028$ & $<.001$ \\
C(sector, Treatment(reference='Manufacturing'))[T.Ethics] & $-0.419$ & $0.036$ & $<.001$ \\
C(sector, Treatment(reference='Manufacturing'))[T.Hospitality] & $-0.177$ & $0.030$ & $<.001$ \\
C(background, Treatment(reference='non-technical'))[T.ai-ml] & $0.337$ & $0.105$ & $0.001$ \\
C(background, Treatment(reference='non-technical'))[T.business] & $0.210$ & $0.064$ & $0.001$ \\
C(background, Treatment(reference='non-technical'))[T.computer-science] & $0.399$ & $0.108$ & $<.001$ \\
C(background, Treatment(reference='non-technical'))[T.data-science] & $0.068$ & $0.121$ & $0.574$ \\
C(background, Treatment(reference='non-technical'))[T.design] & $0.015$ & $0.140$ & $0.917$ \\
C(background, Treatment(reference='non-technical'))[T.engineering] & $-0.062$ & $0.110$ & $0.576$ \\
C(background, Treatment(reference='non-technical'))[T.operations] & $0.260$ & $0.084$ & $0.002$ \\
C(background, Treatment(reference='non-technical'))[T.other] & $0.135$ & $0.073$ & $0.066$ \\
C(background, Treatment(reference='non-technical'))[T.research] & $0.022$ & $0.094$ & $0.816$ \\
C(background, Treatment(reference='non-technical'))[T.software-dev] & $0.256$ & $0.100$ & $0.010$ \\
C(age\_group)[T.25-34] & $-0.044$ & $0.064$ & $0.492$ \\
C(age\_group)[T.35-44] & $-0.094$ & $0.067$ & $0.162$ \\
C(age\_group)[T.45-54] & $-0.134$ & $0.077$ & $0.085$ \\
C(age\_group)[T.55-64] & $-0.136$ & $0.099$ & $0.169$ \\
C(age\_group)[T.65+] & $0.002$ & $0.197$ & $0.993$ \\
C(education\_level)[T.bachelors] & $0.014$ & $0.109$ & $0.899$ \\
C(education\_level)[T.high-school] & $0.015$ & $0.128$ & $0.904$ \\
C(education\_level)[T.masters] & $0.037$ & $0.114$ & $0.745$ \\
C(education\_level)[T.other] & $0.444$ & $0.256$ & $0.083$ \\
C(education\_level)[T.phd] & $-0.037$ & $0.163$ & $0.823$ \\
C(education\_level)[T.professional] & $-0.174$ & $0.178$ & $0.328$ \\
C(education\_level)[T.some-college] & $-0.064$ & $0.113$ & $0.574$ \\
C(gender)[T.male] & $0.074$ & $0.046$ & $0.104$ \\
C(gender)[T.non-binary] & $-0.326$ & $0.153$ & $0.034$ \\
C(gender)[T.prefer-not-to-say] & $-0.417$ & $0.196$ & $0.033$ \\
significance & $0.512$ & $0.042$ & $<.001$ \\
complexity & $-0.033$ & $0.043$ & $0.437$ \\
significance:complexity & $-0.032$ & $0.011$ & $0.005$ \\
\bottomrule
\end{tabular}
\rowcolors{2}{white}{white}

%% file: scr_supplement/S5_pooled_OLS_meta.tex
\rowcolors{2}{white}{rowblue}
\begin{tabular}{lr}
\toprule
Statistic & Value \\
\midrule
$n$ & 1.52e+04 \\
$R^2$ & 0.1606 \\
Adjusted $R^2$ & 0.1588 \\
$n$ clusters & 982 \\
\bottomrule
\end{tabular}
\rowcolors{2}{white}{white}

%% file: scr_supplement/S6_icc.tex
\rowcolors{2}{white}{rowblue}
\begin{tabular}{lr}
\toprule
Quantity & Value \\
\midrule
ICC (readiness) & 0.3248 \\
\bottomrule
\end{tabular}
\rowcolors{2}{white}{white}

%% file: scr_supplement/S6_application_residuals.tex
\rowcolors{2}{white}{rowblue}
\begin{tabular}{lrrr}
\toprule
Application family & Mean residual & SE & $n$ \\
\midrule
Manufacturing & 0.156 & 0.014 & 5244 \\
Biomedical & 0.013 & 0.022 & 2604 \\
Hospitality & -0.030 & 0.019 & 3621 \\
Cyber Security & -0.178 & 0.020 & 2843 \\
Ethics & -0.267 & 0.039 & 888 \\
\bottomrule
\end{tabular}
\rowcolors{2}{white}{white}

%% file: scr_supplement/S7_full_adjusted_coefs.tex
\rowcolors{2}{white}{rowblue}
\begin{tabular}{lrrr}
\toprule
Term & Coef. & SE & $p$ \\
\midrule
Intercept & $1.619$ & $0.195$ & $<.001$ \\
C(sector, Treatment(reference='Manufacturing'))[T.Biomedical] & $-0.106$ & $0.027$ & $<.001$ \\
C(sector, Treatment(reference='Manufacturing'))[T.Cyber Security] & $-0.305$ & $0.028$ & $<.001$ \\
C(sector, Treatment(reference='Manufacturing'))[T.Ethics] & $-0.385$ & $0.036$ & $<.001$ \\
C(sector, Treatment(reference='Manufacturing'))[T.Hospitality] & $-0.239$ & $0.029$ & $<.001$ \\
C(background, Treatment(reference='non-technical'))[T.ai-ml] & $0.310$ & $0.100$ & $0.002$ \\
C(background, Treatment(reference='non-technical'))[T.business] & $0.188$ & $0.063$ & $0.003$ \\
C(background, Treatment(reference='non-technical'))[T.computer-science] & $0.349$ & $0.108$ & $0.001$ \\
C(background, Treatment(reference='non-technical'))[T.data-science] & $0.033$ & $0.116$ & $0.775$ \\
C(background, Treatment(reference='non-technical'))[T.design] & $-0.015$ & $0.137$ & $0.910$ \\
C(background, Treatment(reference='non-technical'))[T.engineering] & $-0.070$ & $0.108$ & $0.518$ \\
C(background, Treatment(reference='non-technical'))[T.operations] & $0.241$ & $0.084$ & $0.004$ \\
C(background, Treatment(reference='non-technical'))[T.other] & $0.122$ & $0.072$ & $0.090$ \\
C(background, Treatment(reference='non-technical'))[T.research] & $0.004$ & $0.094$ & $0.964$ \\
C(background, Treatment(reference='non-technical'))[T.software-dev] & $0.255$ & $0.098$ & $0.009$ \\
C(age\_group)[T.25-34] & $-0.064$ & $0.064$ & $0.312$ \\
C(age\_group)[T.35-44] & $-0.117$ & $0.066$ & $0.075$ \\
C(age\_group)[T.45-54] & $-0.160$ & $0.075$ & $0.034$ \\
C(age\_group)[T.55-64] & $-0.179$ & $0.098$ & $0.066$ \\
C(age\_group)[T.65+] & $-0.034$ & $0.180$ & $0.851$ \\
C(education\_level)[T.bachelors] & $-0.021$ & $0.111$ & $0.851$ \\
C(education\_level)[T.high-school] & $-0.006$ & $0.129$ & $0.966$ \\
C(education\_level)[T.masters] & $0.004$ & $0.115$ & $0.970$ \\
C(education\_level)[T.other] & $0.319$ & $0.231$ & $0.166$ \\
C(education\_level)[T.phd] & $-0.062$ & $0.165$ & $0.709$ \\
C(education\_level)[T.professional] & $-0.208$ & $0.180$ & $0.247$ \\
C(education\_level)[T.some-college] & $-0.082$ & $0.115$ & $0.475$ \\
C(gender)[T.male] & $0.069$ & $0.045$ & $0.126$ \\
C(gender)[T.non-binary] & $-0.288$ & $0.162$ & $0.076$ \\
C(gender)[T.prefer-not-to-say] & $-0.484$ & $0.164$ & $0.003$ \\
significance\_w & $0.342$ & $0.015$ & $<.001$ \\
complexity\_w & $-0.205$ & $0.015$ & $<.001$ \\
significance\_w:complexity\_w & $-0.031$ & $0.011$ & $0.004$ \\
significance\_pm & $0.491$ & $0.046$ & $<.001$ \\
complexity\_pm & $-0.052$ & $0.049$ & $0.290$ \\
\bottomrule
\end{tabular}
\rowcolors{2}{white}{white}

%% file: scr_supplement/S8_background_complexity_slopes.tex
\rowcolors{2}{white}{rowblue}
\begin{tabular}{lrrr}
\toprule
Background & $C^{w}\rightarrow R$ & 95\% CI & $p$ \\
\midrule
engineering & $-0.348$ & $[-0.481,-0.214]$ & $<.001$ \\
design & $-0.300$ & $[-0.461,-0.139]$ & $<.001$ \\
research & $-0.299$ & $[-0.386,-0.213]$ & $<.001$ \\
software dev & $-0.214$ & $[-0.370,-0.058]$ & $0.007$ \\
data science & $-0.206$ & $[-0.322,-0.089]$ & $<.001$ \\
operations & $-0.201$ & $[-0.323,-0.079]$ & $0.001$ \\
non technical & $-0.194$ & $[-0.254,-0.134]$ & $<.001$ \\
other & $-0.186$ & $[-0.280,-0.091]$ & $<.001$ \\
ai ml & $-0.144$ & $[-0.329,0.041]$ & $0.127$ \\
business & $-0.144$ & $[-0.216,-0.072]$ & $<.001$ \\
computer science & $-0.094$ & $[-0.250,0.063]$ & $0.242$ \\
\bottomrule
\end{tabular}
\rowcolors{2}{white}{white}

%% file: scr_supplement/S9_ordered_logit.tex
\rowcolors{2}{white}{rowblue}
\begin{tabular}{lrrrr}
\toprule
Term & Coef. & SE & $p$ & OR \\
\midrule
significance\_w & $0.620$ & $0.017$ & $<.001$ & $1.858$ \\
complexity\_w & $-0.335$ & $0.018$ & $<.001$ & $0.715$ \\
significance\_pm & $0.901$ & $0.033$ & $<.001$ & $2.461$ \\
complexity\_pm & $-0.090$ & $0.032$ & $0.005$ & $0.914$ \\
1/2 & $0.622$ & $0.106$ & $<.001$ & $1.863$ \\
2/3 & $0.357$ & $0.018$ & $<.001$ & $1.429$ \\
3/4 & $0.461$ & $0.012$ & $<.001$ & $1.585$ \\
4/5 & $0.330$ & $0.016$ & $<.001$ & $1.392$ \\
\bottomrule
\end{tabular}
\rowcolors{2}{white}{white}

%% file: scr_supplement/S10_modelB_comparison.tex
\rowcolors{2}{white}{rowblue}
\begin{tabular}{lrrrrrr}
\toprule
Subset & $n$ & $R^2$ B0 & $R^2$ B1 & $\Delta R^2$ & Conf.$^w$ & Trust$^w$ \\
\midrule
Applied pooled & 11469 & 0.208 & 0.434 & 0.227 & $0.199$ & $0.155$ \\
Biomedical & 2604 & 0.157 & 0.433 & 0.276 & $0.173$ & $0.229$ \\
Manufacturing & 5244 & 0.158 & 0.350 & 0.192 & $0.169$ & $0.104$ \\
Hospitality & 3621 & 0.251 & 0.488 & 0.237 & $0.233$ & $0.151$ \\
\bottomrule
\end{tabular}
\rowcolors{2}{white}{white}

%% file: scr_supplement/S10b_modelB_coefs.tex
\rowcolors{2}{white}{rowblue}
\begin{tabular}{llrrr}
\toprule
Subset & Term & Coef. & SE & $p$ \\
\midrule
Applied pooled & confidence\_w & $0.199$ & $0.013$ & $<.001$ \\
Applied pooled & trust\_w & $0.155$ & $0.013$ & $<.001$ \\
Applied pooled & familiarity\_w & $0.096$ & $0.010$ & $<.001$ \\
Applied pooled & convenience\_w & $0.046$ & $0.011$ & $<.001$ \\
Applied pooled & satisfaction\_w & $0.130$ & $0.013$ & $<.001$ \\
Biomedical & confidence\_w & $0.173$ & $0.031$ & $<.001$ \\
Biomedical & trust\_w & $0.229$ & $0.030$ & $<.001$ \\
Biomedical & familiarity\_w & $0.085$ & $0.024$ & $<.001$ \\
Biomedical & convenience\_w & $0.012$ & $0.029$ & $0.664$ \\
Biomedical & satisfaction\_w & $0.144$ & $0.027$ & $<.001$ \\
Manufacturing & confidence\_w & $0.169$ & $0.021$ & $<.001$ \\
Manufacturing & trust\_w & $0.104$ & $0.020$ & $<.001$ \\
Manufacturing & familiarity\_w & $0.094$ & $0.015$ & $<.001$ \\
Manufacturing & convenience\_w & $0.067$ & $0.018$ & $<.001$ \\
Manufacturing & satisfaction\_w & $0.121$ & $0.020$ & $<.001$ \\
Hospitality & confidence\_w & $0.233$ & $0.023$ & $<.001$ \\
Hospitality & trust\_w & $0.151$ & $0.025$ & $<.001$ \\
Hospitality & familiarity\_w & $0.099$ & $0.019$ & $<.001$ \\
Hospitality & convenience\_w & $0.040$ & $0.020$ & $0.045$ \\
Hospitality & satisfaction\_w & $0.123$ & $0.023$ & $<.001$ \\
\bottomrule
\end{tabular}
\rowcolors{2}{white}{white}

%% file: scr_supplement/S10c_trust_conf_contrast.tex
\rowcolors{2}{white}{rowblue}
\begin{tabular}{lrrrrr}
\toprule
Challenge family & Trust$^{w}$ & Confidence$^{w}$ & $\Delta$ (trust$-$conf) & SE & $p$ \\
\midrule
Biomedical ($n=2604$) & 0.226 & 0.174 & $+0.052$ & 0.051 & 0.308 \\
Manufacturing ($n=5244$) & 0.104 & 0.170 & $-0.066$ & 0.032 & 0.039 \\
Hospitality ($n=3621$) & 0.154 & 0.229 & $-0.075$ & 0.038 & 0.048 \\
\bottomrule
\end{tabular}
\rowcolors{2}{white}{white}

%% file: scr_supplement/S11_freetext_sentiment.tex
\rowcolors{2}{white}{rowblue}
\begin{tabular}{lrr}
\toprule
Configuration & Mean sentiment & $n$ \\
\midrule
HighS\_HighR & 0.181 & 316 \\
HighS\_LowR & -0.059 & 198 \\
\bottomrule
\end{tabular}
\rowcolors{2}{white}{white}

%% file: scr_supplement/S11_freetext_themes.tex
\rowcolors{2}{white}{rowblue}
\begin{tabular}{llrr}
\toprule
Configuration & Theme & Count & \% \\
\midrule
HighS\_HighR & privacy\_security & 28 & 11.7 \\
HighS\_HighR & human\_oversight & 61 & 25.5 \\
HighS\_HighR & job\_displacement & 15 & 6.3 \\
HighS\_HighR & readiness\_training & 24 & 10.0 \\
HighS\_HighR & efficiency\_benefit & 39 & 16.3 \\
HighS\_HighR & trust\_reliability & 39 & 16.3 \\
HighS\_HighR & ethics\_fairness & 9 & 3.8 \\
HighS\_HighR & human\_experience & 24 & 10.0 \\
HighS\_LowR & privacy\_security & 24 & 18.0 \\
HighS\_LowR & readiness\_training & 13 & 9.8 \\
HighS\_LowR & human\_experience & 22 & 16.5 \\
HighS\_LowR & human\_oversight & 29 & 21.8 \\
HighS\_LowR & efficiency\_benefit & 10 & 7.5 \\
HighS\_LowR & trust\_reliability & 18 & 13.5 \\
HighS\_LowR & job\_displacement & 8 & 6.0 \\
HighS\_LowR & ethics\_fairness & 9 & 6.8 \\
\bottomrule
\end{tabular}
\rowcolors{2}{white}{white}

%% file: scr_supplement/S12_card_dispersion.tex
\rowcolors{2}{white}{rowblue}
\begin{tabular}{ll r rrr rrr rrr}
\toprule
 &  &  & \multicolumn{3}{c}{Significance} & \multicolumn{3}{c}{Complexity} & \multicolumn{3}{c}{Readiness} \\
\cmidrule(lr){4-6}\cmidrule(lr){7-9}\cmidrule(lr){10-12}
Family & Challenge & $n$ & Mean & SD & IQR & Mean & SD & IQR & Mean & SD & IQR \\
\midrule
Ethics & Ethical awareness & 888 & 4.30 & 0.94 & 1.00 & 3.94 & 1.08 & 2.00 & 3.04 & 1.17 & 2.00 \\
Cyber Security & Cyber incident experience & 918 & 4.18 & 0.93 & 1.00 & 3.90 & 0.98 & 2.00 & 3.14 & 1.12 & 2.00 \\
Cyber Security & Cyber risk preparedness & 949 & 4.25 & 0.87 & 1.00 & 4.03 & 0.93 & 2.00 & 3.05 & 1.11 & 2.00 \\
Cyber Security & Security awareness & 976 & 4.16 & 0.88 & 1.00 & 3.91 & 0.96 & 2.00 & 3.07 & 1.06 & 2.00 \\
Biomedical & Sample handling & 867 & 4.08 & 1.07 & 2.00 & 3.90 & 1.11 & 2.00 & 3.19 & 1.21 & 2.00 \\
Biomedical & Diagnostics & 870 & 4.35 & 0.94 & 1.00 & 4.21 & 1.01 & 1.00 & 3.28 & 1.17 & 1.00 \\
Biomedical & Sample analysis & 867 & 4.10 & 1.03 & 1.00 & 3.94 & 1.06 & 2.00 & 3.31 & 1.15 & 1.00 \\
Manufacturing & Foundation AI support & 872 & 3.94 & 1.00 & 2.00 & 3.54 & 1.14 & 1.00 & 3.55 & 1.10 & 1.00 \\
Manufacturing & Catalyst optimisation & 881 & 3.76 & 1.03 & 2.00 & 3.83 & 1.03 & 2.00 & 3.10 & 1.08 & 2.00 \\
Manufacturing & Battery QC & 877 & 3.84 & 1.05 & 2.00 & 3.74 & 1.03 & 2.00 & 3.26 & 1.07 & 1.00 \\
Manufacturing & Robot precision & 868 & 3.89 & 0.98 & 2.00 & 3.77 & 1.01 & 2.00 & 3.39 & 1.05 & 1.00 \\
Manufacturing & Human-centred manufacturing & 874 & 3.93 & 1.00 & 2.00 & 3.78 & 1.02 & 2.00 & 3.30 & 1.03 & 1.00 \\
Manufacturing & Manufacturing modernisation & 872 & 3.92 & 0.96 & 2.00 & 3.79 & 1.00 & 2.00 & 3.36 & 1.03 & 1.00 \\
Hospitality & Humanoid housekeepers & 907 & 3.17 & 1.23 & 2.00 & 3.37 & 1.25 & 1.00 & 2.89 & 1.26 & 2.00 \\
Hospitality & Robot chefs & 905 & 2.98 & 1.25 & 2.00 & 3.56 & 1.20 & 2.00 & 2.55 & 1.22 & 1.00 \\
Hospitality & Robot spa therapists & 898 & 2.46 & 1.25 & 2.00 & 3.27 & 1.28 & 2.00 & 2.36 & 1.21 & 2.00 \\
Hospitality & Smart travel & 911 & 3.37 & 1.20 & 1.00 & 2.98 & 1.17 & 2.00 & 3.64 & 1.04 & 1.00 \\
\bottomrule
\end{tabular}
\rowcolors{2}{white}{white}